\documentclass[reprint,aps,graphicx,pre,showpacs]{revtex4-1}
\usepackage{amsmath,amssymb,amsfonts}     

\usepackage{graphicx,bm}
  \usepackage{paralist}
  \usepackage{epstopdf}
  \usepackage{graphics} 
 \usepackage[colorlinks=true]{hyperref}

\renewcommand{\theequation}{\arabic{section}.\arabic{equation}}

\usepackage{float}
\usepackage{epsfig}
\usepackage{esint}
\usepackage{chemfig}

\usepackage{bm,braket}

\renewcommand{\theequation}{\arabic{section}.\arabic{equation}}
\newcommand{\calH}{{\mathcal H}}

\newcommand{\calU}{{\mathcal U}}

\newcommand{\calG}{{\mathcal G}}

\newcommand{\calP}{{\mathcal P}}
\newcommand{\calQ}{{\mathcal Q}}
\newcommand{\calS}{{\mathcal S}}

\newcommand{\R}{{\mathbb R}}
\renewcommand{\L}{{\mathbb L}}
\newcommand{\X}{\mathbf{X}}
\renewcommand{\P}{\mathbb{P}}

\newcommand{\x}{\mathbf{x}}

\newcommand{\J}{\widetilde{J}}
\newcommand{\y}{\mathbf{y}}
\newcommand{\e}{{\mathrm e}}

\newcommand{\n}{\mathbf n}

\newcommand{\calT}{{\mathfrak T}}
\newcommand{\calF}{{\mathcal F}}
\renewcommand{\P}{\mathbb P}
\newcommand{\p}{\widetilde{p}}

\newcommand{\f}{\widetilde{f}}
\newcommand{\q}{\widetilde{q}}

\newcommand{\wrho}{\widetilde{\rho}}
\newcommand{\wphi}{\widetilde{\phi}}

\renewcommand{\S}{\widetilde{S}}

\begin{document}

 \title{Random search with stochastic resetting: when finding the target is not enough}

\author{Paul C. Bressloff}
\address{Department of Mathematics, Imperial College London, London SW7 2AZ, UK}

\begin{abstract} 
In this paper we consider a random search process with stochastic resetting and a partially accessible target $\calU$. That is, when the searcher finds the target by attaching to its surface $\partial \calU$ it does not have immediate access to the resources within the target interior.
After a random waiting time, the searcher either gains access to the resources within or detaches and continues its search process. We also assume that the searcher requires an alternating sequence of periods of bulk diffusion interspersed with local surface interactions before being able to attach to the surface. The attachment, detachment and target entry events are the analogs of adsorption, desorption and absorption of a particle by a partially reactive surface in physical chemistry. In applications to animal foraging, the resources could represent food or shelter while resetting corresponds to an animal returning to its home base. We begin by considering a Brownian particle on the half-line with a partially accessible target at the origin $x=0$.
We calculate the non-equilibrium stationary state (NESS) in the case of reversible adsorption and obtain the corresponding first passage time (FPT) density for absorption when adsorption is only partially reversible. We then reformulate the stochastic process in terms of a pair of renewal equations that relate the probability density and FPT density for absorption in terms of the corresponding quantities for irreversible adsorption. The renewal equations allow us to incorporate non-Markovian models of absorption and desorption by taking the waiting time density for the duration of a bound state to be non-exponential. They also provide a useful decomposition of quantities such as the mean FPT (MFPT) in terms of the number of desorption events and the statistics of the waiting time density. Finally, we consider various extensions of the theory, including higher-dimensional search processes and an encounter-based model of absorption. The latter assumes that absorption only occurs when the total time the particle is attached to the target exceeds a randomly generated threshold, irrespective of the number of return visits.

\end{abstract}

\maketitle
\section{Introduction}

\begin{figure*}[t!]
\centering
\includegraphics[width=15cm]{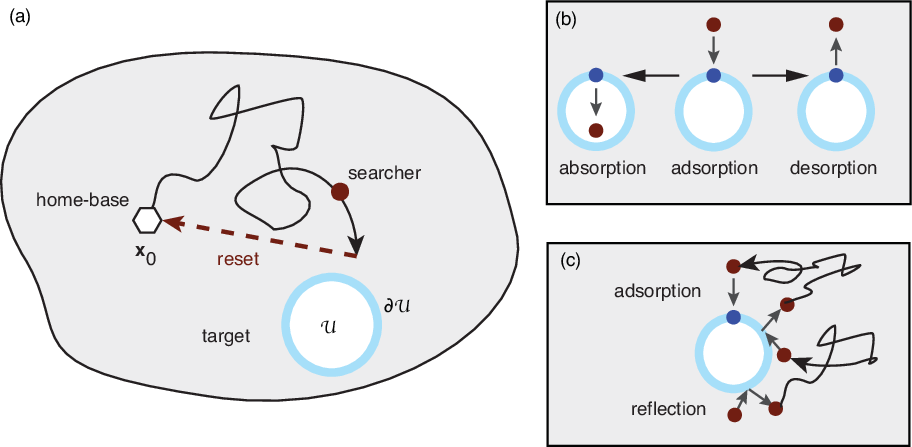} 
\caption{Search for a partially accessible target. (a) A foraging animal diffusively searches for a hidden target $\calU$ that contains resources such as food or shelter. Prior to finding the target, the forager may return to its home base at $\x_0$ in order to rest. (b) Adsorption of the searcher at a point on $\partial \calU$ does not give it immediate access to the resources within $\calU$. After some random waiting time attached to the surface, the searcher either succeeds in entering the interior $\calU$ (absorption) or detaches from the target to continue its search for an accessible target (desorption). (c) There is a non-zero probability that the searcher reflects off the target surface rather than attaching (partial adsorption).}
\label{fig1}
\end{figure*}

Stochastic resetting is a mechanism that returns a system to its initial state at a random sequence of times that is usually taken to be a Poisson process with constant rate $r$. Following the original example of a Brownian particle that instantaneously resets to its initial position $\x_0\in \R^d$ \cite{Evans11a,Evans11b,Evans14}, there have been a wide range of generalisations at both the single and multiple particle  levels, see the reviews \cite{Evans20,Nagar23} and references therein. These include both modifications in the underlying stochastic dynamics in the absence of resetting and modifications in the resetting protocol itself. Within the context of animal foraging, stochastic resetting has been proposed as a mechanism for reducing the expected time to find a hidden target $\calU\subset \R^d$ within some large search domain. The target represents a local region of resources such as food or shelter, and resetting mimics the observed tendency for an animal to return to its home base in order to rest or resupply \cite{Pal20}, see Fig. \ref{fig1}(a). Target detection is said to occur when the particle first reaches a point on the target boundary $\partial \calU$, which can be implemented by taking $\partial \calU$ to be a totally absorbing surface. The total search time is given by the first passage time (FPT) $T=\inf\{t>0,\ \X(t)\in \partial \calU\}$, where $\X(t)\in \R^d$ is the position of the searcher at time $t$. If the searcher fails to find the target then $T=\infty$. One of the limitations of a purely diffusive process as a stochastic search mechanism is that the mean first passage time (MFPT) $\langle T\rangle$ diverges as the size of the search domain goes to infinity. On the other hand, the introduction of a stochastic resetting protocol can support a finite MFPT that has a unique minimum as a function of the resetting rate \cite{Evans11a,Evans11b,Evans14}. 

In spite of the extensive number of studies of search processes with stochastic resetting, there has been relatively little attention paid to the nature of target-searcher interactions beyond some work on partially adsorbing surfaces \cite{Evans13,Bressloff22b,Benk22}. In this paper we modify the standard formulation of random search with resetting by considering a partially accessible target, see Fig. \ref{fig1}(b). First, we assume that when the searcher finds the target by attaching to its surface $\partial \calU$, it does not have immediate access to the resources within the target interior $\calU$. Instead,  it spends a random waiting time $\tau$ with associated density $\phi(\tau)$ attached to the surface, after which it either gains access to the resources or detaches and continues its search process. We will refer to these two alternatives as {\em absorption} and {\em desorption}, respectively. Second, we assume that the searcher requires an alternating sequence of periods of bulk diffusion interspersed with local surface interactions before being able to attach to the surface. In other words, the reactive surface $\partial \calU$ is {\em partially adsorbing}. We consider two alternative models for what happens immediately after desorption. One assumes that the searcher immediately returns to its home base before continuing the search process, whereas the other takes the search process to continue from the point on the surface $\partial \calU$ where the searcher detaches. As a further simplification, we do not include any memory effects regarding the location of the target. This is motivated by the notion that the searcher does not know how many targets are located within the search domain. Therefore it continues to explore the domain using a random search strategy even though it may repeatedly return to the same target.

The structure of the paper is as follows. In Sect. II we introduce the theory by considering the simple example of a Brownian particle on the half-line with a partially accessible target at the origin $x=0$ and instantaneous resetting to its initial position $x_0$. (In this paper we neglect the effects of finite return times \cite{Pal19,Pal19a,Mendez19,Bodrova20,Pal20,Bressloff20} and refractory periods \cite{Evans19a,Mendez19a} associated with random sojourns at home base.) We treat the boundary as partially adsorbing in the sense that whenever the particle hits the boundary there is a non-zero probability that it is reflected rather than adsorbed. Taking the adsorption rate to be a constant $\kappa_0$, the boundary condition of the single-particle diffusion equation at $x=0$ is of the Robin type. Once the particle is adsorbed, it unbinds (desorbs) at a constant rate $\gamma_0$ or is permanently removed (absorbed) at a rate $\overline{\gamma}_0$. If $\gamma_0=0$ and $\overline{\gamma}_0\rightarrow \infty$ then we recover the standard example of irreversible adsorption. On the other hand, if $\gamma_0>0$ and $\overline{\gamma}_0=0$, then we have an example of reversible adsorption.
We first calculate the non-equilibrium stationary state (NESS) in the case of reversible adsorption and determine how this is affected by desorption. We then derive the FPT density for absorption when adsorption is only partially reversible ($\gamma_0,\overline{\gamma}_0>0$). 

In Sect. III we reformulate the diffusion equation as a pair of renewal equations that relate the probability density and FPT density for absorption in terms of the corresponding quantities for irreversible adsorption. The renewal equations effectively sew together successive rounds of adsorption and desorption prior to the final absorption event. One advantage of the renewal approach is that it is straightforward to incorporate non-Markovian model of absorption and desorption by taking the waiting time density for the duration of a bound state to be non-exponential. (The Markovian exponential case is equivalent to taking constant rates of desorption and absorption.) The renewal equations also provide an efficient way of determining the FPT for absorption if the corresponding FPT for adsorption is already known. In particular, quantities of interest such as the MFPT can be expressed in terms of the number of desorption events and the statistics of the waiting time density. We illustrate this by calculating the first two moments of the FPT density for absorption, and determining how they depend on various parameters including the resetting rate, the mean and variance of the waiting time density, and the splitting probabilities of desorption versus absorption. In Sect. IV we construct a higher-dimensional version of the renewal equations and solve them in the particular case of a spherically symmetric target. A more general analysis based on spectral theory is presented in Appendix A. Finally, in Sect. V, we consider an encounter-based mechanism for generating non-Markovian absorption, whereby the probability of absorption depends on the total amount of time the particle is attached to the target surface irrespective of the number of return visits.

We end this introduction by highlighting the fact that reversible surface adsorption-desorption processes (without resetting) have been been studied for many years in physical chemistry \cite{Baret68,Adam87,Adam87a,Franses95,Passerone96,Reuveni23} and have many features in common with 
reversible recombination–dissociation reactions based on Smoluchowski theory \cite{Agmon84,Agmon89,Agmon90,Agmon93,Gopich99,Kim99,Tachiya13,Prustel13,Grebenkov19}. Encounter-based models of diffusion-mediated non-Markovian surface adsorption have also been extended to include the effects of desorption, both in the reversible case \cite{Grebenkov23} and most recently in the partially reversible case \cite{Bressloff25}. These latter studies use renewal equations analogous to the ones considered in the current paper.  

\section{Diffusion on the half-line with a partially accessible target}

\begin{figure}[b!]
\centering
\includegraphics[width=8.5cm]{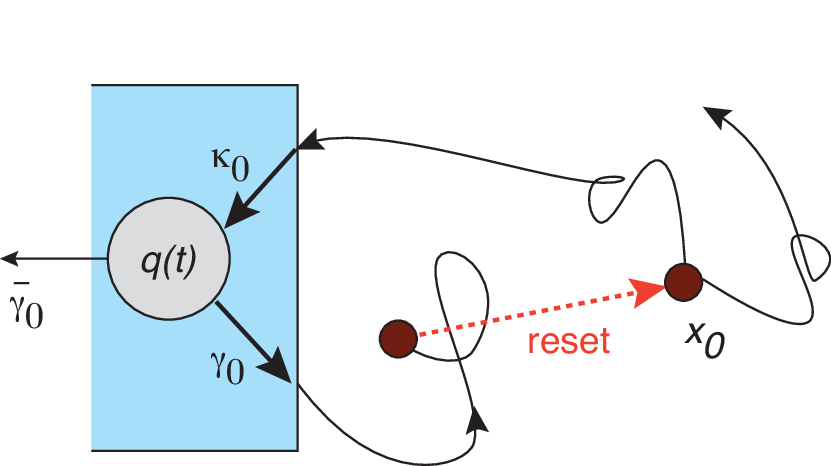} 
\caption{A diffusing particle in the half-line $[0,\infty)$ with a partially reactive boundary at $x=0$. The particle binds to the surface boundary (adsorbs) at a rate $\kappa_0$ and then either desorbs at a rate $\gamma_0$ or is permanently absorbed at a rate $\overline{\gamma}_0$. The probability of being in the bound state is $q(t)$. While diffusing in the bulk domain, the particle instantaneously resets to its initial position at a rate $r$.}
\label{fig2}
\end{figure}

Consider a particle undergoing diffusion with instantaneous stochastic resetting on the half-line $x\in [0,\infty)$ with a partially reactive boundary at $x=0$, see Fig. \ref{fig2}. Whenever the particle hits the boundary at $x=0$, it either reflects or enters a bound state at a constant rate $\kappa_{0}$. We denote the probability that the particle is in the bound state at time $t$ by $q(t)$. The bound particle then unbinds at a rate $ \gamma_0$ or is permanently removed at a rate $\overline{\gamma}_0$. Whilst diffusing in the bulk domain $(0,\infty)$, the particle resets to its initial position $x_0>0$ according to a Poisson process with rate $r$. The probability density $\rho(x,t|x_0)$ evolves according to the equation
\begin{subequations}
\label{1D}
\begin{align}
 \frac{\partial \rho(x,t|x_0)}{\partial t}&=D\frac{\partial^2 \rho(x,t|x_0)}{\partial x^2} -r\rho(x,t|x_0)\nonumber \\
 &\quad +r \calS(x_0,t)\delta(x-x_0),\quad x>0,\\
D\frac{\partial \rho(0,t|x_0)}{\partial x}&=\kappa_0\rho(0,t|x_0) -\gamma_0 q(x_0,t), 
\end{align}
with
\begin{equation}
\frac{dq(x_0,t) }{dt}=\kappa_0\rho(0,t|x_0) - (\gamma_0+\overline{\gamma}_0) q(x_0,t).
\end{equation}
\end{subequations}
The initial conditions are $\rho(x,0|x_0)=\delta(x-x_0)$ and $q(x_0,0)=0$.
We have also introduced the survival probability that the particle is freely diffusing at time $t$,
\begin{equation}
\label{Sr}
\calS(x_0,t)=\int_0^{\infty} \rho(x,t|x_0)dx.
\end{equation}

In the limit $\gamma_0\rightarrow 0$ (no desorption), we recover the classical Robin boundary condition for 1D diffusion with resetting, which has been analysed elsewhere \cite{Evans13,Bressloff22b}. On the other hand, if $\gamma_0>0$ then either adsorption is reversible ($\overline{\gamma}_0=0$) or partially reversible ($\overline{\gamma}_0>0$). In the latter case, the particle is ultimately absorbed and removed from the system. This motivates the introduction of a second survival probability
\begin{equation}
\label{Qr}
\calQ(x_0,t)=1-\overline{\gamma}_0\int_0^t q(\tau)d\tau,
\end{equation}
which represents the probability that the particle has not yet been absorbed (irrespective of whether it is freely diffusing or bound at $x=0$.)
Differentiating Eqs. (\ref{Sr}) and (\ref{Qr}) with respect to time $t$ and using Eqs. (\ref{1D}a)-(\ref{1D}c) shows that
\begin{equation}
\label{dSr}
\frac{d\calS(x_0,t)}{dt}=-D\frac{\partial \rho(0,t|x_0)}{\partial x},
\end{equation}
and
\begin{equation}
\label{dQr}
\frac{d\calQ(x_0,t)}{dt}\equiv \frac{d\calS(x_0,t)}{dt}+\frac{dq(t)}{dt}=- \overline{\gamma}_0  q(t).
\end{equation}
Note that $-d\calS/dt$ equals the net flux into the boundary from the bulk and $-d\calQ/dt$ is the absorption flux (rate of killing). If $\overline{\gamma}_0=0$, then Eq. (\ref{dQr}) ensures conservation of probability such that
\begin{equation}
\int_0^{\infty}\rho(x,t|x_0)dx+q(t)=1
\end{equation}
for all $t$. Note that the model described by Eqs. (\ref{1D}) assumes that after desorption the particle continues the search process from $x=0$ rather than $x=x_0$. The latter case is more complicated to formulate using partial differential equations (PDEs). The converse holds for the equivalent renewal equations, see Sect. III.

\begin{figure}[t!]
\centering
\includegraphics[width=8cm]{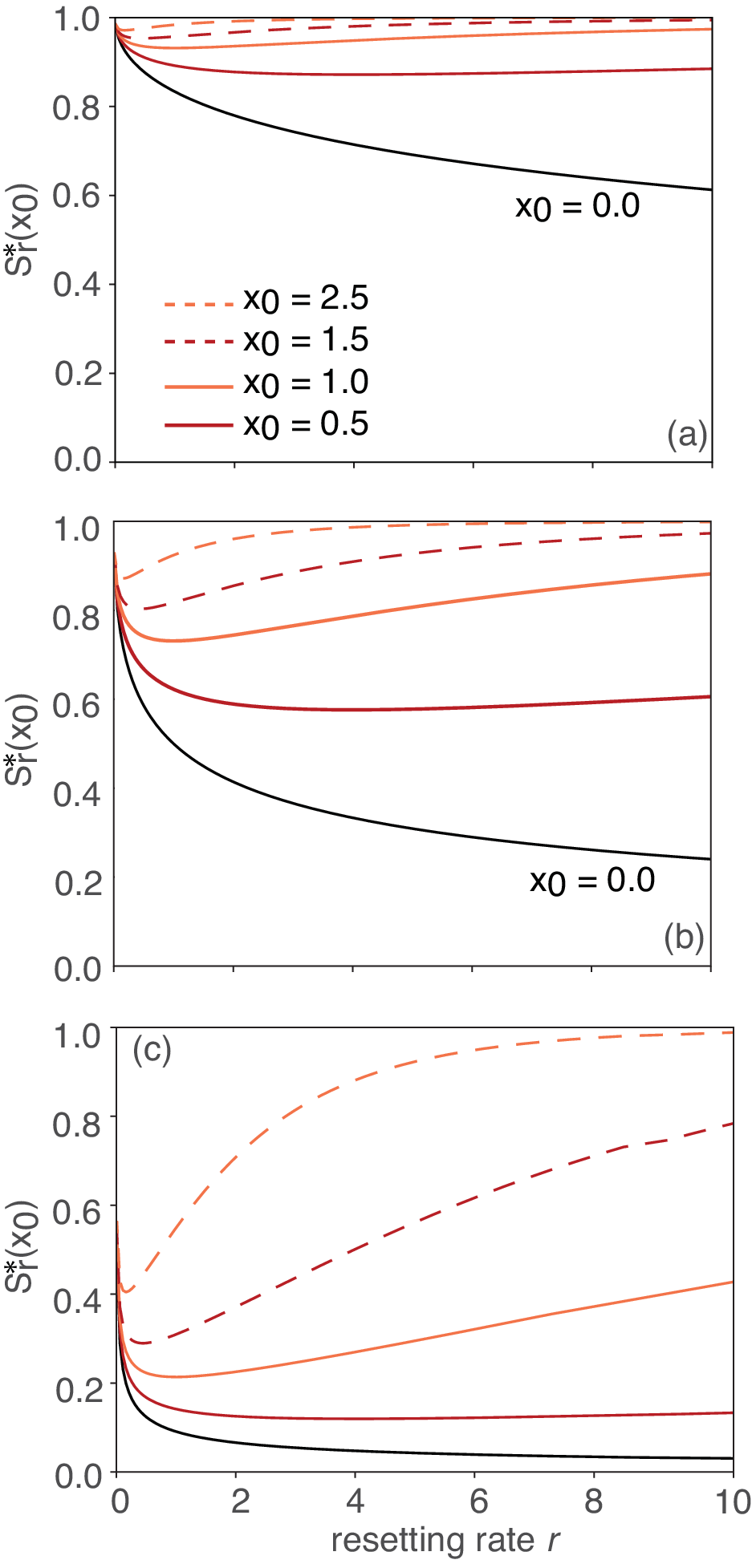} 
\caption{Steady-state survival probability $ \calS^*(x_0)$ for a diffusing particle in the half-line $[0,\infty)$ with a reversibly adsorbing boundary at $x=0$ and stochastic resetting at a rate $r$. (a) $\kappa_0/\gamma_0=0.2$. (b) $\kappa_0/\gamma_0=1$. (c) $\kappa_0/\gamma_0=10$. Diffusivity is $D=1$.}
\label{fig3}
\end{figure}

Laplace transforming equations (\ref{1D}a)--(\ref{1D}c) gives
\begin{subequations}
\label{1DLT}
\begin{align}
 &D\frac{\partial^2\wrho(x,s|x_0)}{\partial x^2} -(r+s)\wrho(x,s|x_0)\nonumber \\
 & =-[1+r \widetilde{\calS}(x_0,s)]\delta(x-x_0),\quad x>0,\\
 &D\frac{\partial \wrho(0,s|x_0)}{\partial x}=\kappa(s)\wrho(0,s|x_0),\\
 &\q(x_0,s)=\frac{\kappa_0}{s+\gamma_0+\overline{\gamma}_0}\wrho(0,s|x_0),
\end{align}
\end{subequations}
with
\begin{equation}
\label{ks}
\kappa(s)=\frac{(s+\overline{\gamma}_0)\kappa_0}{s+\gamma_0+\overline{\gamma}_0}.
\end{equation}
The general bounded solution of equation (\ref{1DLT}a) is of the form
\begin{equation}
\label{gen1D}
\wrho(x,s|x_0)=C(s)\e^{-\alpha x}+[1+r\widetilde{\calS}(x_0,s)]G(x,\alpha|x_0),
\end{equation}
where $\alpha=\sqrt{[r+s]/D}$.
The first term on the right-hand side of Eq. (\ref{gen1D}) is the solution to the homogeneous version of Eq. (\ref{1D}a) and $G$ is the Dirichlet Green's function of the 1D modified Helmholtz equation on the half-line:
 \begin{eqnarray}
\label{GG1D}
 G(x, \alpha|x_0) = \frac{1}{2D\alpha }\left [\e^{-\alpha |x-x_0|}-\e^{-\alpha |x+x_0|}\right ].
 \end{eqnarray}
  The unknown coefficient $C(s)$ is determined by imposing the boundary condition (\ref{1DLT}b):
 \begin{eqnarray}
 C(s)=\frac{1+r\widetilde{\calS}(x_0,s)}{\kappa(s) +\alpha D}\e^{-\alpha x_0}.
 \end{eqnarray}
 Hence, the full solution of the Laplace transformed probability density with resetting is
  \begin{align}
   \wrho(x, s|x_0) &= \bigg  [1+r\widetilde{\calS}(x_0,s) \bigg ]\nonumber \\
   &\quad \times \left (\frac{\e^{-\alpha (x+x_0)}}{\kappa(s) +\alpha D}+G(x,\alpha|x_0)\right )
\label{p1D}
\end{align}
for $x>0$.
 Finally, Laplace transforming Eq. (\ref{Sr}) and using (\ref{p1D}) shows that
 \begin{align}
  \widetilde{\calS}(x_0,s)&=\int_0^{\infty}\wrho(x,s|x_0)dx\nonumber \\
 &=[1+r\widetilde{\calS}(x_0,s)]\widetilde{\calS}_0(x_0,r,s),
\label{SSr}
 \end{align}
 where
 \begin{equation}
 \label{SS0}
\widetilde{\calS}_0(x_0,r,s)=\frac{1-\e^{-\alpha x_0}}{r+s}+\frac{\e^{-\alpha x_0}}{r+s+\alpha\kappa(s)}.
 \end{equation}
  Rearranging Eq. (\ref{SSr}) then determines the Laplace transformed survival probability with resetting in terms of the function $\widetilde{\calS}(x_0,r,s)$:
 \begin{equation}
 \label{Sren}
\widetilde{\calS}(x_0,s)=\frac{\widetilde{\calS}_0(x_0,r,s)}{1-r\widetilde{\calS}_0(x_0,r,s)}.
 \end{equation}
 It is important to note that although $\widetilde{\calS}_0(x_0,r=0,s)=\widetilde{\calS}_0(x_0,s)$ is the Laplace transform of the survival probability without resetting, we cannot set $\widetilde{\calS}_0(x_0,r,s)=\widetilde{\calS}_0(x_0,r+s)$. This is a consequence of the fact that   adsorption/desorption leads to the $\kappa(s)$-dependent term in Eq. (\ref{SS0}). Hence, Eq. (\ref{Sren}) differs from the standard result obtained from the renewal theory of stochastic processes with resetting \cite{Evans20}.
 
 \subsection{Reversible adsorption ($\overline{\gamma}_0=0$)}
\begin{figure}[t!]
\centering
\includegraphics[width=8cm]{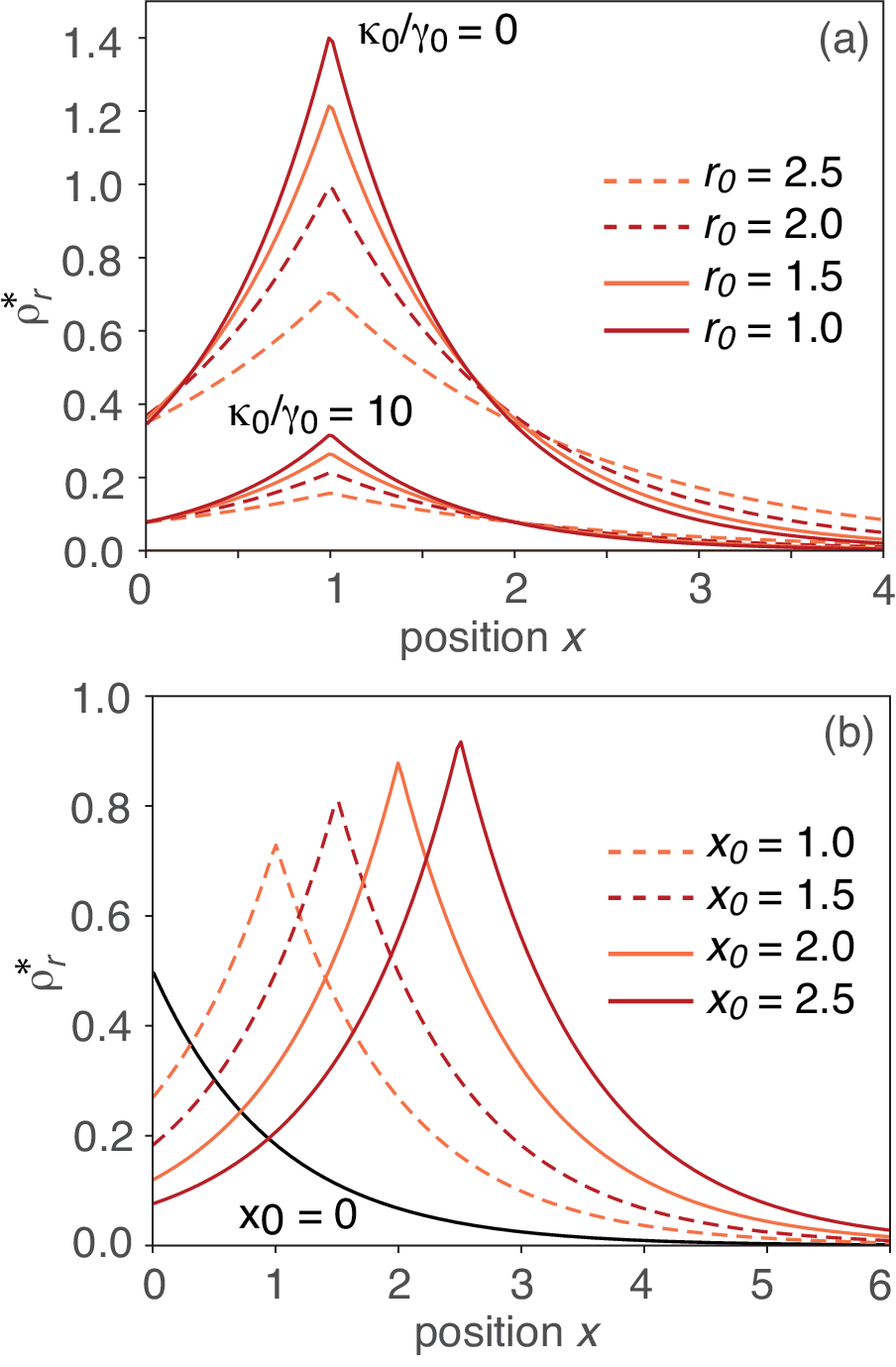} 
\caption{NESS for a diffusing particle in the half-line $[0,\infty)$ with a reversibly adsorbing boundary at $x=0$ and stochastic resetting at a rate $r$, see Eq. (\ref{NESS1D}). (a) Plots of $\rho^*(x|x_0)$ as a function of $x$ for various $r,\kappa_0/\gamma_0=0,10$ and $x_0=1$. (b)  Corresponding plots for various $x_0$ with $r=1$ and $\kappa_0=1$. Diffusivity is $D=1$.}
\label{fig4}
\end{figure}

 For $\overline{\gamma}_0=0$, we have $\calQ(x_0,t)=1$ for all $t\geq 0$ so that there exists a nonequilibrium stationary state (NESS). In order to determine the NESS we first need to find the steady-state survival probability $\calS^*(x_0)$. Eq. (\ref{SSr}) implies that
 \begin{align}
 \calS^*(x_0)&:=\lim_{t\rightarrow \infty} \calS(x_0,t)=\lim_{s\rightarrow 0}s\widetilde{\calS}(x_0,s)\\
 & =\lim_{s\rightarrow 0}\frac{s\widetilde{\calS}(x_0,r,s)}{1-r\widetilde{\calS}(x_0,r,s)}.\nonumber 
 \end{align}
 Moreover, Eq. (\ref{ks}) becomes $\kappa(s)=\kappa_0s/(s+\gamma_0)$ so that, after some algebra,
  \begin{align}
 \calS^*(x_0)=  & \frac{1}{1+\sqrt{r/D}\e^{-\sqrt{r/D}x_0}(\kappa_0/\gamma_0)}.
 \end{align}
If $\kappa_0=0$ then the boundary is totally reflecting and $\calS^*(x_0)=1$ as expected. On the other hand, in the limit $\gamma_0\rightarrow 0$ there is no desorption and $\calS^*(x_0)=0$. Plots of $\calS^*(x_0)$ as a function of the resetting rate $r$ are shown in Fig. \ref{fig3}. It can be seen that for $x_0>0$, the survival probability is a unimodal function of $r$ and $\lim_{r\rightarrow \infty}\calS^*(x_0)=1$, since the particle never has time to reach the boundary. Given the stationary state $\calS^*(x_0)$, the corresponding NESS is obtained from Eq. (\ref{p1D}) and takes the form
  \begin{align}
\label{NESS1D}
&\rho^*(x|x_0)  = \widetilde{\calS}^*(x_0) 
    \sqrt{\frac{r}{D}} \e^{-\sqrt{r/D} |x-x_0|}  .
\end{align}
That is, $\rho^*(x|x_0) $ is equal to the NESS for a totally reflecting boundary scaled by the factor $\widetilde{\calS}^*(x_0) $. The effect of this scale factor is illustrated in Fig. \ref{fig4}. 

 \subsection{Partially reversible adsorption ($\overline{\gamma}_0>0$)}
 If we now include the effects of absorption by taking $\overline{\gamma}_0>0$ then a non-trivial NESS no longer exists. The quantity of interest becomes the FPT density for absorption, which is defined according to   
\begin{equation}
\calF(x_0,t)=-\frac{d\calQ(x_0,t)}{dt}=\overline{\gamma}_0  q(t).
\end{equation}
Combining Eqs. (\ref{1DLT}c), (\ref{p1D}) and (\ref{Sren}) gives
\begin{align}
\widetilde{\calF}(x_0,s)=\frac{\kappa_0\overline{\gamma}_0}{s+\gamma_0+\overline{\gamma}_0} \frac{1}{1-r\widetilde{\calS}(x_0,r,s)}\frac{\e^{-\alpha x_0}}{\kappa(s) +\alpha D}.
\label{qrs}
\end{align}
The Laplace transform of the FPT density is the moment generating function, as will be explored further in Sect. III within the context of the renewal approach.

 \setcounter{equation}{0}
 
 \section{Renewal equations}
 
 Recently we have developed a general probabilistic framework for analyzing stochastic processes with partially reversible adsorbing boundaries using renewal theory \cite{Bressloff25}, which builds upon a previous study of the reversible case \cite{Grebenkov23}. Here we reinterpret the results obtained in Sect. II within the renewal theory framework. The renewal equations relate the densities $\rho(x,t|x_0)$ and $\calF(x_0,t)$ in the presence of desorption and absorption (partially reversible adsorption) to the corresponding quantities without desorption (irreversible adsorption). One of the useful features of the renewal approach is that it is straightforward to incorporate non-Markovian desorption and absorption processes. Following Ref. \cite{Bressloff25}, we will assume for simplicity that when the particle is adsorbed, it remains bound for a random time $\tau$ generated from a waiting time density $\phi(\tau)$. The particle then either desorbs with a splitting probability $\pi_d$ or is permanently absorbed with probability $\pi_b=1-\pi_d$. In the exponential case
 \begin{equation}
 \label{ephi}
 \phi(\tau)=\gamma\e^{-\gamma \tau},\quad \gamma =\gamma_0+\overline{\gamma}_0,
 \end{equation}
 with the associated splitting probabilities
 \begin{equation}
 \label{split}
 \pi_d=\frac{\gamma_0}{\gamma_0+\overline{\gamma}_0},\quad \pi_b=\frac{\overline{\gamma}_0}{\gamma_0+\overline{\gamma}_0},
 \end{equation}
 we recover the Markovian scheme considered in Sect. II.
 
 \begin{figure*}[t!]
\centering
\includegraphics[width=18cm]{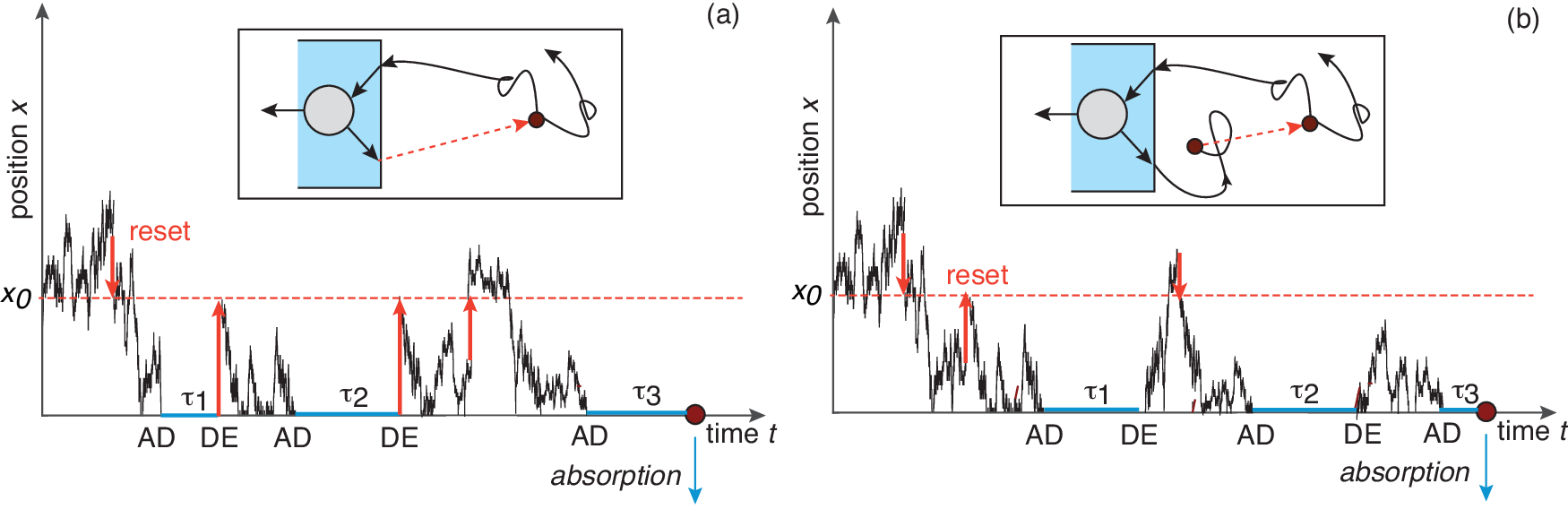} 
\caption{Example trajectories of a Brownian particle with instantaneous stochastic resetting on the half-line consisting of two rounds of adsorption (AD) and desorption (DE) prior to being absorbed. The sequence of waiting times in the bound state are given by $\{\tau_1,\tau_2,\tau_3\}$. (a) The particle resets to $x_0$ immediately after each desorption event. (b) The particle continues searching from $x=0$ after each desorption event. }
\label{fig5}
\end{figure*}
 
 First, suppose that the particle resets to $x_0$ immediately after desorption. The renewal equations are
\begin{subequations}
\begin{align}
\label{ren10}
  &\overline{\rho}(x,t|x_0)=p(x,t|x_0)\\
  & +\pi_d\int_0^td\tau' \int_{\tau'}^t d\tau\,  \overline{\rho}(x,t-\tau|x_0)  \phi_r(\tau-\tau') f(x_0,\tau'),\nonumber \\
   \label{ren20}
  & \overline{\calF}(x_0,t)=\pi_b\int_0^td\tau  {\phi}(t-\tau) f(x_0,\tau)\\
  &+\pi_d \int_0^td\tau' \int_{\tau'}^t d\tau\, \overline{\calF}(x_0,t-\tau) \phi(\tau-\tau') f(x_0,\tau').\nonumber 
 \end{align}
 \end{subequations}
(Here $p$ and $f$ are the solutions in the case of irreversible adsorption. We also distinguish $\overline{\rho}$ and $\overline{\calF}$ from the solutions obtained by continuing the search process from $x=0$ following desorption, see below.)
The first term on the right-hand side of the renewal equation (\ref{ren10}) for the density represents the contribution from all sample paths that start at $x_0$ and have not been adsorbed over the interval $[0,t]$. On the other hand, the second term represents all sample paths starting from $x_0$ that are first adsorbed at a time $\tau' $ with probability $f(x_0,\tau')d\tau'$, remain in the bound state until desorbing at time $\tau$ with probability $\pi_d \phi(\tau-\tau')d\tau$, after which the particle may bind an arbitrary number of times before reaching $x$ at time $t$. Turning to the renewal equation (\ref{ren2}) for the FPT  $\calF(x_0,t)$, the first term on the right-hand side represents all sample paths that are first adsorbed at time $\tau$ and are subsequently absorbed at time $t$ without desorbing, which occurs with probability $\pi_b{\phi}(t-\tau)f(x_0,\tau)d\tau dt$. In a complementary fashion, the second term sums over all sample paths that are first adsorbed at time $\tau'$, desorb at time $\tau$ and are ultimately absorbed at time $t$ following an arbitrary number of additional adsorption events. One can view the renewal equations as sewing together successive rounds of adsorption and desorption prior to the final absorption event, as illustrated in Fig. \ref{fig5}(a).

If the particle continues the search process from $x=0$ after desorption, see Fig. \ref{fig5}(b), then we need to generalize the renewal equations by allowing the reset point $x_r$ to be distinct from the initial position $x_0$. We indicate this by adding a subscript $r$ to all corresponding quantities.
The resulting renewal equations now take the form
\begin{subequations}
\begin{align}
\label{ren1}
  &\rho_r(x,t|x_0)=p_r(x,t|x_0)\\
  & +\pi_d\int_0^td\tau' \int_{\tau'}^t d\tau\,  \rho_r(x,t-\tau|0)  \phi_r(\tau-\tau') f_r(x_0,\tau'),\nonumber \\
   \label{ren2}
  & \calF_r(x_0,t)=\pi_b\int_0^td\tau  {\phi}(t-\tau) f_r(x_0,\tau)\\
  &+\pi_d \int_0^td\tau' \int_{\tau'}^t d\tau\, \calF_r(0,t-\tau) \phi(\tau-\tau') f_r(x_0,\tau').\nonumber 
 \end{align}
 \end{subequations}
 Here $\rho_r(x,t|z)$ and $ \calF_r(z,t)$, $z=0,x_0$, denote the densities for trajectories that start at $x=z$ and reset to $x=x_r$. A similar convention will be used for the corresponding densities for irreversible adsorption. Having solved the renewal equations we can then set $x_r=x_0$ as in Sect. II.

\subsection{Solution in Laplace space}
Both versions of the renewal equations can be solved using Laplace transforms and the convolution theorem. First, Laplace transforming Eqs. (\ref{ren10}) and (\ref{ren20}) and rearranging shows that
\begin{subequations}
\begin{align}
\widetilde{\overline{\rho}}(x,s|x_0)&=\frac{\p(x,s|x_0)}{1-\pi_d \wphi(s) \f(x_0,s)}
\label{LTren10}
\end{align}
and
\begin{align}
 \widetilde{\overline{\calF}}(x_0,s)
 &=\frac{\pi_b \wphi(s) \f(x_0,s)}{1-\pi_d \wphi(s) \f(x_0,s)}.
  \label{LTren20}
 \end{align}
 \end{subequations}
On the other hand, Laplace transforming Eq. (\ref{ren1}) gives
\begin{align}
\wrho_r(x,s|x_0)&=\p_r(x,s|x_0) \\
&\quad + \wrho_r(x,s|0)\pi_d \wphi(s) \f_r(x_0,s).\nonumber
 \end{align}
 Setting $x_0=0$ and rearranging shows that
 \begin{equation}
 \wrho_r(x,s|0)=\frac{\p_r(x,s|0)}{1-\pi_d \wphi(s) \f_r(0,s)}
 \end{equation}
 and, hence,
 \begin{align}
\wrho_r(x,s|x_0)&=\p_r(x,s|x_0)\nonumber \\
&\quad +\frac{\pi_d \wphi(s) \f_r(x_0,s)}{1-\pi_d \wphi(s) \f_r(0,s)}\p_r(x,s|0).
\label{LTren1}
 \end{align}
Similarly, Laplace transforming the second renewal equation (\ref{ren2}) gives
 \begin{align}
& \widetilde{\calF}_r(x_0,s)\\
&=\wphi(s)\bigg [\pi_b \f_r(x_0,s)+\pi_d \widetilde{\calF}_r(0,s) \f_r(x_0,s)\bigg ].\nonumber 
 \end{align}
 Setting $x_0=0$ and rearranging implies that
 \begin{equation}
  \widetilde{\calF}_r(0,s)=\frac{\pi_b\wphi(s) \f_r(0,s)}{1-\pi_d\wphi(s) \f_r(0,s)},
  \end{equation}
 and thus
  \begin{align}
 \widetilde{\calF}_r(x_0,s)
 &=\frac{\pi_b \wphi(s) \f_r(x_0,s)}{1-\pi_d \wphi(s) \f_r(0,s)}.
  \label{LTren2}
 \end{align}
Substituting the relation $\f_r(x_0,s)=\kappa_0 \p_r(0,s|x_0)$ into the right-hand side of (\ref{LTren2}) implies that 
  \begin{align}
 \widetilde{\calF}_r(x_0,s) &=\kappa_0\pi_b   \wphi(s)\wrho_r(0,s|x_0).
\label{LTren2a}
 \end{align}
 In the case of the exponential waiting time density (\ref{ephi}) and splitting probabilities (\ref{split}), we have
 \begin{equation}
 \wphi(s)=\frac{\gamma_0+\overline{\gamma}_0}{s+\gamma_0+\overline{\gamma}_0},
 \end{equation}
 and
  \begin{align}
 \widetilde{\calF}_r(x_0,s)
 &=\frac{\kappa_0 \overline{\gamma}_0}{s+\gamma_0+\overline{\gamma}_0} \wrho_r(0,s|x_0).
 \end{align}
 Hence, we recover the boundary conditions (\ref{1DLT}c) after setting $x_r=x_0$.

It remains to calculate the various expressions in the case of irreversible adsorption. First note that $\p_r(x,s|x_0)$ satisfies the boundary value problem (BVP)
\begin{subequations}
\label{1DLT0}
\begin{align}
 &D\frac{\partial^2\p_r(x,s|x_0)}{\partial x^2} -(r+s)\wrho_r(x,s|x_0)\nonumber \\
 & =-\delta(x-x_0)-r \S_r(x_0,s)]\delta(x-x_r),\quad x>0,\\
 &D\frac{\partial \p_r(0,s|x_0)}{\partial x}=\kappa_0\p_r(0,s|x_0).
 \end{align}
\end{subequations}
Proceeding along similar lines to the derivation of Eq. (\ref{p1D}) we find that
  \begin{align}
 \label{p1D0}
   \p_r(x, s|x_0) &=
 \p_0(x,r+ s|x_0) \\
&\quad +r \S_r(x_0,s)\p_0(x,r+ s|x_r),\nonumber
\end{align}
and
\begin{align}
\S_r(x_0,s)&\equiv \int_0^{\infty}\p_r(x,s|x_0)dx\nonumber \\
&=\frac{\S_0(x_0,r+s)}{1-r\S_0(x_r,r+s)},
 \label{Sren0}
 \end{align}
 where
\begin{equation}
\p_0(x,r+ s|x_0)=\frac{\e^{-\alpha (x+x_0)}}{\kappa_0 +\alpha D}+G(x,\alpha|x_0)
\end{equation}
and
 \begin{equation}
 \label{SS00}
\S_0(x_0,r+s)=\frac{1-\e^{-\alpha x_0}}{r+s}+\frac{\e^{-\alpha x_0}}{r+s+\alpha \kappa_0}.
 \end{equation}
The densities $\p_0(x,s|x_0)$ and $\S_0(x_0,s)$ are the Laplace transforms of the probability density and survival probability without resetting nor desorption ($r=0$ and $\gamma_0=0$). The corresponding FPT density is
 \begin{align}
 \label{ff00}
 \f_0(x_0,s)\equiv 1-s\S_0(x_0,s)=\frac{\kappa_0\sqrt{s/D}}{s+\sqrt{s/D}\kappa_0}\e^{-\sqrt{s/D}x_0}.
 \end{align}
 Finally, the density $\f_r(x_0,s)$ is given by
 \begin{align}
 \f_r(x_0,s)&=1-s\S_r(x_0,s)\\
 &=\frac{ r\f_0(x_r,r+s)+s\f_0(x_0,r+s) }{s+r\f_0(x_r,r+s)}.\nonumber 
  \end{align}
  We thus find that
\begin{align}
 \label{ivf}
\f_r(x_0,s)
&=\frac{\alpha \kappa_0 \left [r\e^{-\alpha x_r}+se^{-\alpha x_0}\right ]}{s(r+s)+\alpha \kappa_0(s+r \e^{-\alpha x_r} )}.
\end{align}

It can be checked that the solutions $\wrho_r(x,s|x_0)$ and $\widetilde{\calF}_r(x_0,s)$ of the renewal equations reduce to the corresponding solutions $\wrho(x,s|x_0)$ and $\widetilde{\calF}(x_0,s)$ of Eqs. (\ref{1D}), after setting $x_r=x_0$ and taking  $\wphi(s)$ to be the exponential distribution, see Eqs. (\ref{ephi}) and (\ref{split}). 
However, the renewal equations are more general since they allow for non-Markovian forms of desorption and absorption in which the waiting time density is non-exponential. They also provide a useful way of decomposing the contributions to quantities such as the MFPT as we now show.

 \subsection{Moments of the FPT density for absorption}

Eq. (\ref{LTren2}) can be used to express the moments of the FPT density $\widetilde{\calF}_r(x_0,s)$ (assuming they exist) in terms of the corresponding moments of $\f_r(x_0,s)$. This follows from the fact that $\f_r(x_0,s)$ and $\widetilde{\calF}_r(x_0,s)$ are moment generating functions:
\begin{subequations}
 \begin{align}
 T_r^{(n)}(x_0):&=\int_0^{\infty}t^n f_r(x_0,t)dt\nonumber \\
 &=\left . \left (-\frac{d}{ds}\right )^n \f_r(x_0,s)\right |_{s=0},\\
 \calT_r^{(n)}(x_0):&=\int_0^{\infty}t^n \calF_r(x_0,t)dt\nonumber \\
 &=\left . \left (-\frac{d}{ds}\right )^n \widetilde{\calF}_r(x_0,s)\right |_{s=0}.
 \end{align}
 \end{subequations}
In particular, $\calT_r=\calT_r^{(1)}$ and $T_r=T_r^{(1)}$ are the MFPTs with and without desorption. Assuming that $\phi(\tau)$ has finite moments,
we substitute the series expansions
 \begin{subequations}
 \label{asym0}
 \begin{align}
  \f_r(x_0s)&\sim 1-sT_r (x_0)+s^2T_r^{(2)}(x_0)/2+O(s^3),\\ \wphi(s)&\sim 1-s\langle \tau\rangle +s^2\langle \tau^2\rangle/2+O(s^3)
  \end{align}
 \end{subequations}
into Eqs. (\ref{LTren2}) and Taylor expand in powers of $s$. Collecting the $O(s)$ terms yields
 \begin{align}
 \calT_r(x_0)=T_r(x_0)+  \langle \tau \rangle+\frac{\pi_d}{\pi_b}\bigg [T_r(0)+ \langle \tau \rangle\bigg ].
\label{MFPT}
 \end{align}
 The corresponding MFPT without desorption is obtained by evaluating $\f_r(x_0,s)$ using Eq (\ref{ivf}): 
\begin{align}
T_r(x_0)=\frac{1}{r}\left (1+\frac{\sqrt{rD}}{\kappa_0}\right )\e^{\sqrt{r/D}x_r}-\frac{1}{r}\e^{\sqrt{r/D}(x_r-x_0)}.
\end{align}
This result was previously derived in Ref. \cite{Evans13}.

 \begin{figure}[t!]
\centering
\includegraphics[width=8cm]{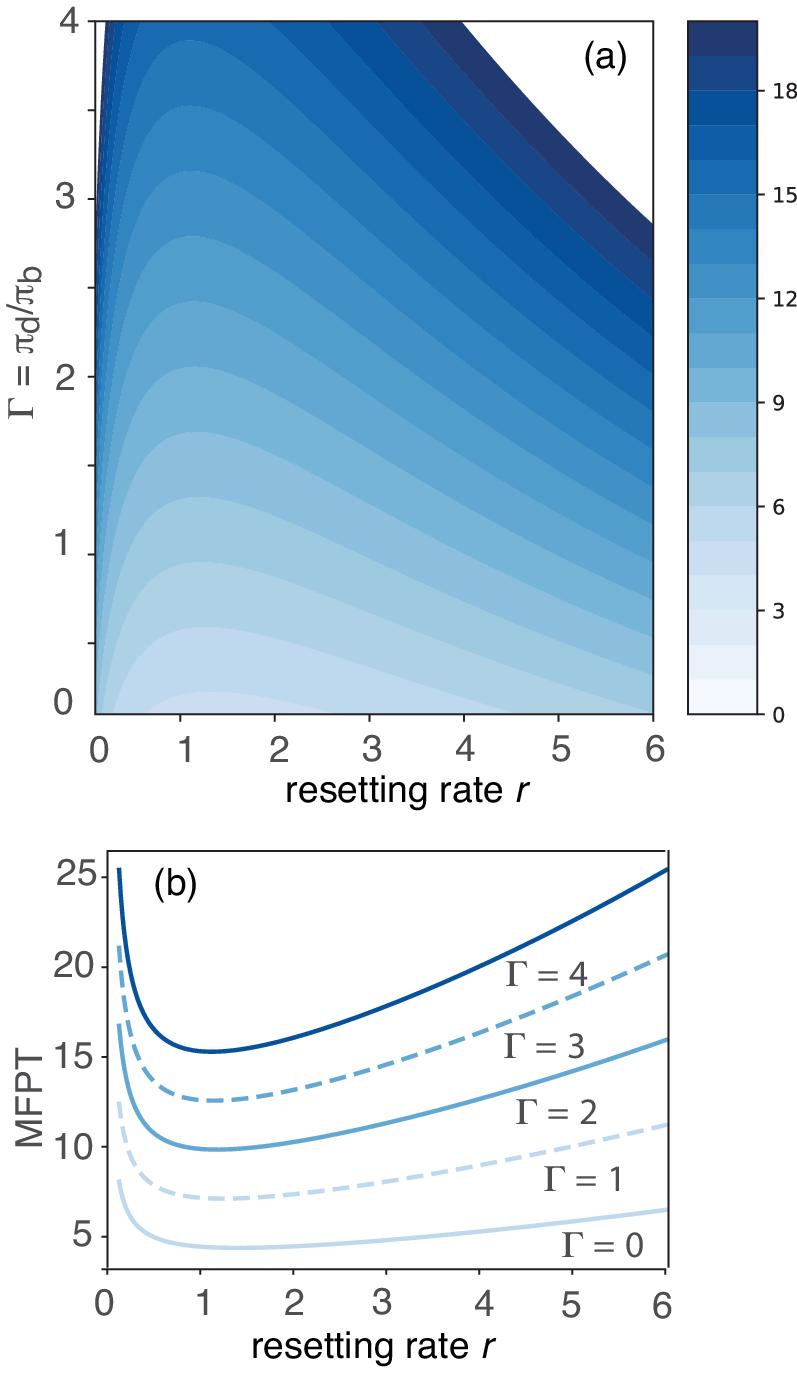} 
\caption{Effect of desorption on the MFPT $\calT_r(x_0)$ for absorption. (a) Contour plot of  $\calT_r(x_0)$ in the $(r,\Gamma)$-plane where $\Gamma=\pi_d/\pi_b$. (b) Plot of $\calT_r(x_0)$ versus $r$ for different values of $\Gamma$. Other parameters are $\langle \tau\rangle =0$, $\kappa_0=1$, $x_0=x_r=1$ and $D=1$.}
\label{fig6}
\end{figure}

Eq.(\ref{MFPT}) has a simple physical interpretation, see also Ref. \cite{Bressloff25}. The first and second terms on the right-hand side are, respectively, the times to be adsorbed for the first time starting from $x_0$ and the mean time to be absorbed following the final adsorption event. The other terms account for the mean time accrued due to desorption. The probability of exactly $n$ desorption events is $p_n=\pi_b\pi_d^n$ with
\begin{equation}
\sum_{n=0}^{\infty}p_n= \pi_b\sum_{n=0}^{\infty}\pi_d^n=\frac{\pi_b}{1-\pi_d}=1.
\end{equation}
The mean number of such excursions is 
\begin{eqnarray}
\overline{n}=\sum_{n=0}^{\infty}np_n=\frac{\pi_b\pi_d}{(1-\pi_d)^2}=\frac{\pi_d}{\pi_b},
\end{eqnarray}
and the mean time between excursions is $T_r(0)+ \langle \tau \rangle$.
As one would expect, the MFPT $ \calT_r(x_0)$ is a monotonically increasing function of $x_0$, $\Gamma\equiv \pi_d/\pi_b$, the mean waiting time $\langle \tau\rangle$ and the mean adsorption time $1/\kappa_0$. It also inherits the classical unimodal dependence on the resetting rate $r$ from $T_r(x_0)$. This is illustrated in Fig. \ref{fig6}, which shows a contour plot of the MFPT in the $(r,\Gamma)$-plane together with plots for fixed $\Gamma$. (For the moment we ignore the effects of waiting times by setting $\langle \tau\rangle =0$.) It can be seen that increasing $\Gamma$ does not simply shift the MFPT curves upwards due to the dependence on both $T_r(x_0)$ and $T_r(0)$. On the other hand, if the searcher returns to home base after each desorption event then we would simply have $\calT_r(x_0)=T_r(x_0)/\pi_b$. (More generally, the FPT moments when the particle immediately resets to $x_0$ after desorption are obtained simply by taking the argument of every term to be $x_0$.)

\begin{figure}[t!]
\centering
\includegraphics[width=8cm]{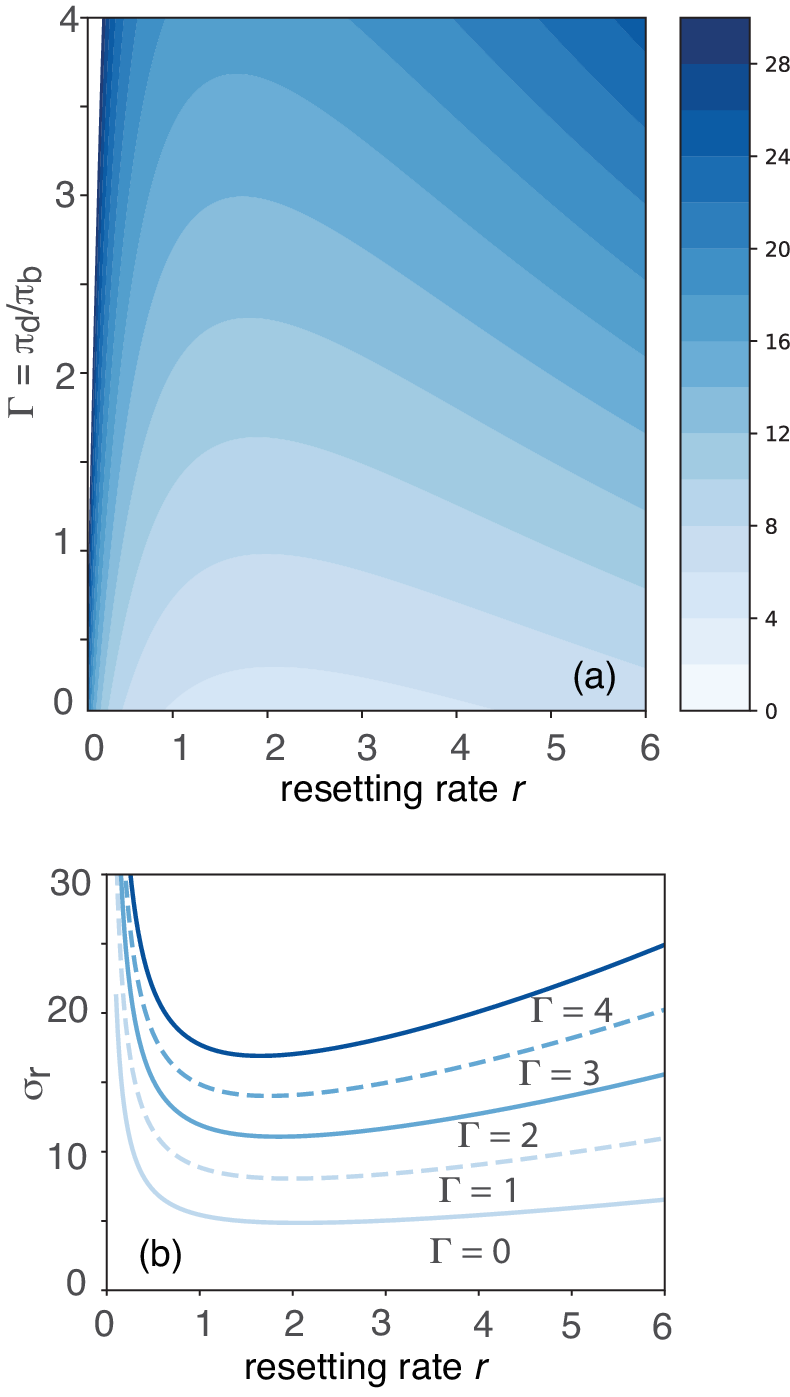} 
\caption{Effect of desorption on the standard deviation $\sigma_r(x_0)$ for absorption. (a) Contour plot of  $\sigma_r(x_0)$ in the $(r,\Gamma)$-plane where $\Gamma=\pi_d/\pi_b$. (b) Plot of $\sigma_r(x_0)$ versus $r$ for different values of $\Gamma$. Other parameters are $\langle \tau\rangle =0=\langle \tau^2\rangle$, $\kappa_0=1$, $x_0=x_r=1$ and $D=1$.}
\label{fig7}
\end{figure}

 \begin{figure}[t!]
\centering
\includegraphics[width=8cm]{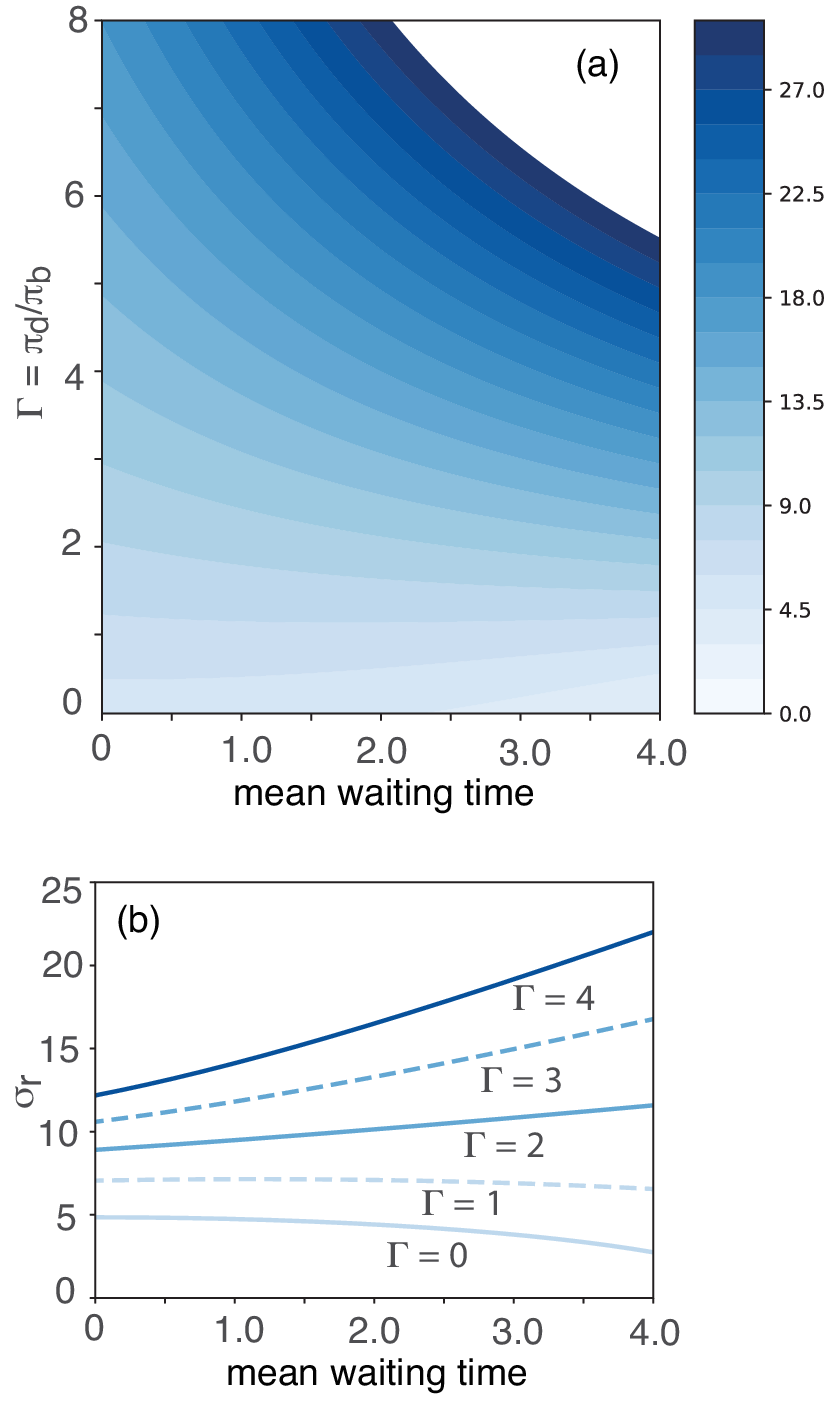} 
\caption{Effect of desorption on the standard deviation $\sigma_r(x_0)$ for absorption. (a) Contour plot of  $\sigma_r(x_0)$ in the $(\langle \tau\rangle,\Gamma)$-plane where $\langle \tau\rangle$ is the mean waiting time and $\Gamma=\pi_d/\pi_b$. (b) Plot of $\sigma_r(x_0)$ versus $\langle \tau\rangle$ for different values of $\Gamma$. Other parameters are $\sqrt{\langle \tau^2\rangle} =4$, $\kappa_0=1$, $x_0=x_r=1$ and $D=1$.}
\label{fig8}
\end{figure}

 \begin{figure}[t!]
\centering
\includegraphics[width=8cm]{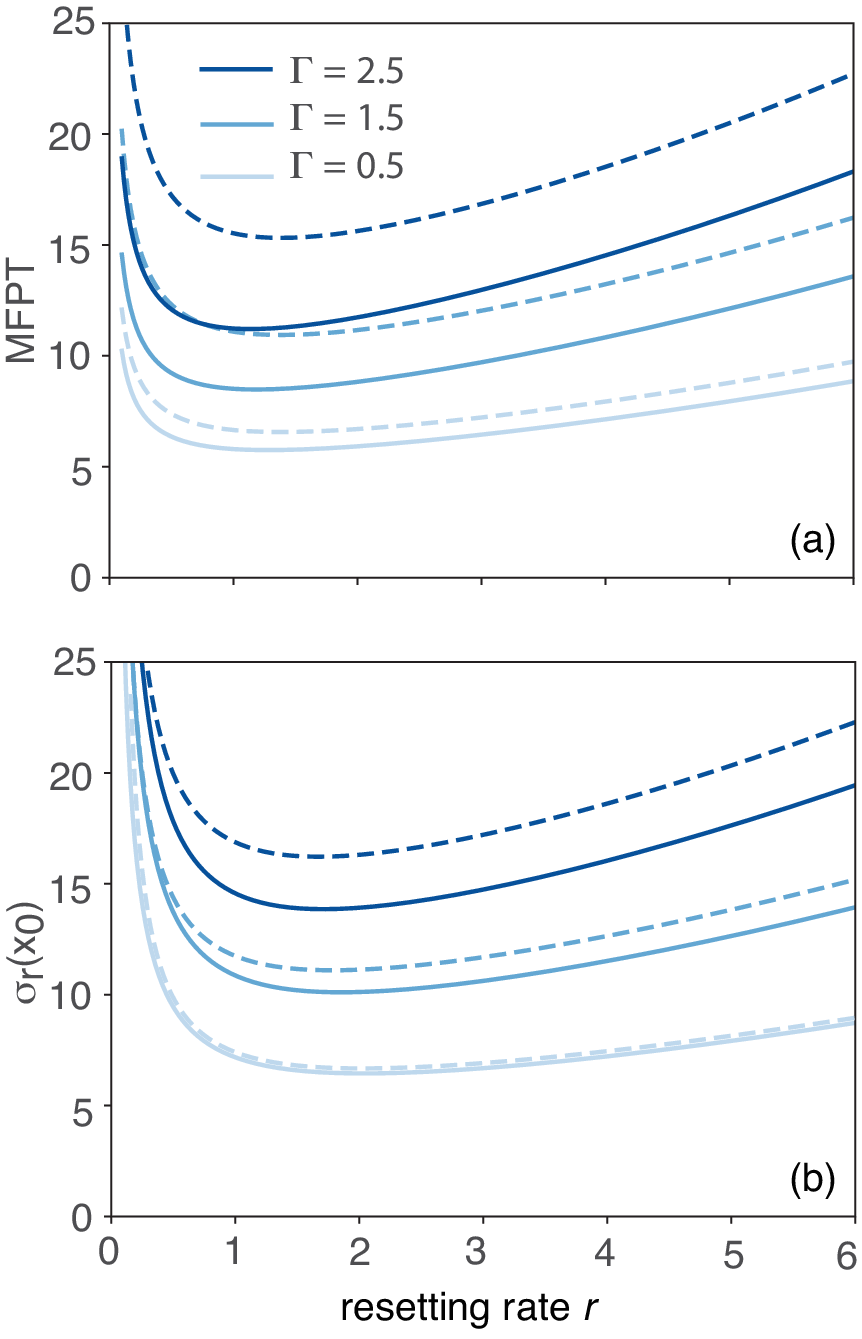} 
\caption{Effect of restart position following desorption on (a) the MFPT $\calT_r(x_0)$ and (b) the standard deviation $\sigma_r(x_0)$ as functions of $r$ and various values of $\Gamma$. Solid (dashed) curves correspond to restarting at $x=0$ ($x=x_0$) following desorption. Other parameters are $\kappa_0=1$, $x_0=x_r=1$, $\langle \tau\rangle =\langle \tau^2\rangle =0$, and $D=1$.}
\label{fig9}
\end{figure}


Higher-order moments have a more complicated dependence on $\Gamma$ and the moments of $\phi(\tau)$. Collecting the $O(s^2)$ terms in the Taylor expansion of (\ref{LTren2}) yields the second-order FPT moment
\begin{align}
  \calT_r^{(2)}(x_0) &=T_r^{(2)}(x_0)+\langle \tau^2\rangle + 2\langle \tau\rangle T_r(x_0)  
  \nonumber \\
  &\quad  +\frac{\pi_d}{\pi_b}\bigg [T_r^{(2)}(0)+\langle \tau^2\rangle + 2\langle \tau\rangle T_r(0)\bigg ]  
  \nonumber \\
 &\quad   +\frac{\pi_d}{\pi_b}\bigg [T_r(x_0)+ \langle \tau \rangle\bigg ]\bigg [T_r(0)+ \langle \tau \rangle\bigg ]  
  \nonumber \\
  &\quad  +2\left (\frac{\pi_d}{\pi_b}\right )^2\bigg [T_r(0)+ \langle \tau \rangle\bigg ]^2,
\label{MFPT2}
 \end{align}
 with $T_r^{(2)}(x_0)$ obtained from Eq. (\ref{ivf}):
  \begin{align}
 & T_r^{(2)}(x_0)\nonumber \\
 &=\frac{1}{r^2}\bigg [1+\sqrt{\frac{r}{D}}(x_r-x_0) \bigg ]\e^{\sqrt{r/D}(x_r-x_0)} \nonumber \\
 &-\frac{1}{r^2}\bigg [1+\frac{2\sqrt{rD}}{\kappa_0}+\frac{\sqrt{rD}}{\kappa_0}\sqrt{\frac{r}{D}}x_r  \bigg ]\e^{\sqrt{r/D}x_r}\nonumber \\
 &+\frac{2}{r^2}\left (1+\frac{\sqrt{rD}}{\kappa_0}\right )^2\e^{2\sqrt{r/D}x_r} .
  \label{Tr2}
  \end{align}
Note that in the absence of desorption ($\Gamma=0$),
 \begin{equation}
\calT_r^{(2)}(x_0)-\calT_r(x_0)^2=T_r^{(2)}(x_0)-T_r(x_0)^2+\langle \tau^2\rangle -\langle\tau\rangle^2.
\end{equation}
That is, the variance of the FPT is the sum of the variances arising from irreversible adsorption and the variance in the waiting time before absorption.
 In Fig. \ref{fig7} we show plots of the standard deviation $\sigma_r(x_0)$ analogous to those of the MFPT in Fig. \ref{fig6}, where
 \begin{equation}
 \sigma_r(x_0)\equiv \sqrt{\calT_r^{(2)}(x_0)-\calT_r(x_0)^2}.
 \end{equation}
 The standard deviation exhibits similar behavior to the MFPT, although the value of $r$ that minimizes $\sigma_r(x_0)$ differs from the optimal value for the MFPT. Finally, in Fig. \ref{fig8}, we show plots of the standard deviation $\sigma_r(x_0)$ as a function of the MFPT $\langle \tau\rangle$ for a fixed second moment $\langle \tau^2\rangle $. In the case of an exponential waiting time density $\phi(\tau)=\gamma \e^{-\gamma \tau}$, fixing $\langle \tau^2\rangle$ also fixes the mean since $\langle \tau^2\rangle=2\langle \tau\rangle^2=2\gamma^2$. However, the renewal equations allow for a non-exponential waiting time density such that $0\leq \langle \tau\rangle \leq \langle \tau^2\rangle$. 
 
 Finally, in Fig. \ref{fig9} we compare the MFPT and standard deviation for the two distinct restart locations  after desorption, namely, $x=0$ as assumed in Figs. \ref{fig6}--\ref{fig8} and $x=x_0$. We find that both $\calT(x_0)$ and $\sigma(x_0)$ increase when the particle returns to home base after desorption and, as might be expected, the size of the increase grows with the likelihood of desorption $\Gamma$.

\setcounter{equation}{0}
\section{Higher-dimensional search for a partially accessible target.}

Let us return to the higher-dimensional configuration shown in Fig. \ref{fig1}(a) for a target surface $\partial \calU$ with  $\calU\subset  \R^d$. The higher-dimensional version of Eqs. (\ref{1D}) take the form
\begin{subequations}
\label{hD}
\begin{align}
 \frac{\partial \rho(\x,t|\x_0)}{\partial t}&=D{\bm \nabla}^2\rho(\x,t|\x_0)  -r\rho(\x,t|\x_0)\nonumber \\
 &\quad +r \calS(\x_0,t)\delta(\x-\x_0),\\
D{\bm \nabla} \rho(\y,t|\x_0)\cdot \n&=\kappa_0\rho(\y,t|\x_0) -\gamma_0 q(\y,t|\x_0), 
\end{align}
with $\x,\x_0\notin \overline{\calU}$, $\y \in \partial \calU$ and
\begin{equation}
\frac{\partial q(\y,t|\x_0) }{\partial t}=\kappa_0\rho(\y,t|\x_0) - (\gamma_0+\overline{\gamma}_0) q(\y,t|\x_0).
\end{equation}
\end{subequations}
Here $\overline{\calU}$ denotes the closure of $\calU$, that is, $\overline{\calU}=\calU\cup \partial \calU$, and $\n$ is the outward unit normal at a point on $\partial \calU$.
The initial conditions are $\rho(\x,0|\x_0)=\delta(\x-\x_0)$ and $q(\y,0|\x_0)=0$ for all $\x \notin \overline{ \calU}$ and $\y \in \partial \calU$.
The survival probability that the particle is freely diffusing at time $t$ is given by
\begin{equation}
\label{Sr2D}
\calS(\x_0,t)=\int_{\R^d\backslash \calU}\rho(\x,t|\x_0)d\x.
\end{equation}

\subsection{Renewal equations} The higher-dimensional versions of the renewal equations (\ref{ren10})--(\ref{ren20}) and (\ref{ren1})--(\ref{ren2}) are, respectively,
\begin{widetext}
\begin{subequations}
\label{2Dren0}
\begin{align}
\overline{ \rho}(\x,t|\x_0)&=p(\x,t|\x_0)+\pi_d\int_0^td\tau' \int_{\tau'}^t d\tau\,  \overline{\rho}(\x,t-\tau|\x_0)\phi(\tau-\tau') f(\x_0,\tau' ),  \\
\overline{ \calF}(\x_0,t)=&\pi_b  \int_0^td\tau {\phi}(t-\tau) f(\x_0,\tau)  
   + \pi_d  \int_0^td\tau' \int_{\tau'}^t d\tau\, \overline{\calF}(\y,t-\tau)\phi(\tau-\tau') f(\x_0,\tau'),
 \end{align}
 \end{subequations}
 and
\begin{subequations}
\label{2Dren}
\begin{align}
 \rho_r(\x,t|\x_0)&=p_r(\x,t|\x_0)+\pi_d\int_{\partial \calU}d\y\int_0^td\tau' \int_{\tau'}^t d\tau\,  \rho_r(\x,t-\tau|\y)\phi(\tau-\tau') J_r(\y,\tau'|\x_0),  \\
 \calF_r(\x_0,t)=&\pi_b\int_{\partial \calU}d\y \int_0^td\tau {\phi}(t-\tau) J_r(\y,\tau|\x_0)  
   + \pi_d \int_{\partial \calU} d\y\int_0^td\tau' \int_{\tau'}^t d\tau\, \calF_r(\y,t-\tau)\phi(\tau-\tau') J_r(\y,\tau'|\x_0).
 \end{align}
 \end{subequations}
 \end{widetext}
Here $p(\x,t|\x_0)$ is the probability density in the absence of desorption and $f(\x_0,t)$ is the corresponding FPT density under the assumption that the particle resets to $\x_0$ as soon as it desorbs. Similarly, $p_r(\x,t|\x_0)$ is the probability density in the absence of desorption and resetting to a general point $\x_r$ and $J_r(\y,t|\x_0)$ is the corresponding adsorption probability flux into the target at $\y\in \partial \calU$. We see a major difference from the 1D case due to the spatially extended nature of the target surface $\partial \calU$. That is, if the particle continues searching from the point $\y\in \partial \calU$ where it desorbs, then it is necessary to include a spatial integral in the renewal equations.

Laplace transforming the renewal Eqs. (\ref{2Dren}) and rearranging shows that
\begin{subequations}
\begin{align}
\widetilde{\overline{\rho}}(\x,s|\x_0)&=\frac{\p(\x,s|\x_0)}{1-\pi_d \wphi(s) \f(\x_0,s)},
\label{LTren100}\\
 \widetilde{\overline{\calF}}(\x_0,s)
 &=\frac{\pi_b \wphi(s) \f(\x_0,s)}{1-\pi_d \wphi(s) \f(\x_0,s)},
  \label{LTren200}
 \end{align}
 \end{subequations}
 which are identical in form to the 1D case. This no longer holds for Eqs. (\ref{2Dren}), which become
\begin{subequations}
\label{2DLT}
 \begin{align}
 \wrho_r(\x,s|\x_0)&=\p_r(\x,s|\x_0)\\
&\quad +\pi_d \wphi(s) \int_{\partial \calU}   \wrho_r(\x,s|\y) \J_r(\y,s|\x_0)d\y,\nonumber \\
\widetilde{\calF}_r(\x_0,s)&= \pi_b\wphi(s) \f_r(\x_0,s) \\
&\quad +\pi_d \wphi(s)  \int_{\partial \calU}  \widetilde{\calF}_r(\y,s) \J_r(\y,s|\x_0)d\y,\nonumber
 \end{align}
 \end{subequations}
 where $ \f_r(\x_0,s)$ is the Laplace transform of the FPT for irreversible adsorption:
 \begin{equation}
 f_r(\x_0,t)=\int_{\partial \calU} J_r(\y,t|\x_0)d\y.
 \end{equation}
  The Laplace transformed probability density $\p_r$ is the solution of the following Robin BVP:
 \begin{subequations}
\begin{align}
&D{\bm \nabla}^2 \p_r(\x,s|\x_0)-(r+s)\p_r(\x,s|\x_0)\\
&=-\delta(\x-\x_0)-r\delta(\x-\x_r)\S_r(\x_0,s),\ \x \notin \overline{\calU}, \nonumber \\
  \J_r(\y,s|\x_0)&\equiv D{\bm \nabla} \p_r(\y,s|\x_0) \cdot \n\nonumber \\
  &=\kappa_0\p_r(\y,s|\x_0) ,\ \y\in \partial \calU .
\end{align}
\end{subequations}
The solution can be expressed as
\begin{align}
\p_r(\x,s|\x_0)&=\p_0(\x,r+s|\x_0) \nonumber \\
&\quad +r\S_r(\x_0,s)\p_0(\x,r+s|\x_r),
\label{peep}
\end{align}
with
 \begin{subequations}
 \label{PlocLT}
\begin{align}
&D{\bm \nabla}^2 \p_0(\x,s|\x_0)-s\p_0(\x,s|\x_0)=-\delta(\x-\x_0),\ \x \notin \overline{\calU},\\
  &\J_0(\y,s|\x_0)\equiv D{\bm \nabla} \p_0(\y,s|\x_0) \cdot \n 
  =\kappa_0\p_0(\y,s|\x_0) ,\ \y\in \partial \calU ,
 \end{align}
\end{subequations}
and
\begin{align}
\S_r(\x_0,s)&\equiv \int_{\R^d\backslash \calU} \p_r(\x,s|\x_0)d\x\nonumber \\
&=\frac{\S_0(\x_0,r+s)}{1-r\S_0(\x_r,r+s)}.
 \label{dSren0}
 \end{align}

One method for handling the spatial integrals is to use  spectral theory. A well known result from the classical theory of partial differential equations is that the solution of a general Robin BVP on a compact surface $\partial \calU$ can be computed in terms of the spectrum of a D-to-N (Dirichlet-to-Neumann) operator \cite{Grebenkov19a}. The basic idea is to replace the Robin boundary condition (\ref{PlocLT}b) by the inhomogeneous Dirichlet condition $\p_0(\y,s|\x_0)=h(\y,s|\x_0)$ for all $\y\in \partial \calU$ and to find the function $h$ for which $\p_0$ is also the solution of the original BVP. This spectral method has recently been applied to the higher-dimensional renewal equations for both reversible adsorption \cite{Grebenkov23} and partially reversible adsorption \cite{Bressloff25} in the absence of stochastic resetting. However, the resulting solutions involve infinite sums that need to be truncated using an appropriate approximation scheme. We develop the corresponding analysis with stochastic resetting in Appendix A. Here we avoid such technicalities by considering a spherical target.

\subsection{Spherically symmetric target.} Let
$\calU=\{\x\in \R^d \, |\, 0\leq  |\x|<R_1\}$ with $ \partial \calU=\{\x\in \R^d\, |\, |\x|=R_1\}$.
Suppose that the spherical target $\partial \calU$ is partially accessible with a constant adsorption rate $\kappa_0$ and waiting time density $\phi(\tau)$ for desorption/absorption. Following \cite{Redner01}, the initial position of the particle is randomly chosen from the surface of the sphere $\calU_0$ of radius $R_0$, $R_1<R_0$. Similarly, we assume that the particle resets to a random point on a sphere $\calU_r$ of radius $R_r$ with $R_1<R_r$. (We will ultimately make the identification $R_0=R_r$.) The search process is then itself spherically symmetric and we can set $\wrho_r=\wrho_r(R,s|R_0)$  etc, in spherical polar coordinates. Eq. (\ref{2DLT}a) thus reduces to the simpler form
 \begin{align}
  \label{sprenewal}
  \widetilde{\rho}_r(R,s|R_0) &= \p_r(R,s|R_0)  \\
  &+\pi_d \wphi(s) \wrho_r(R,s|R_1)\f_r(R_0,s),\nonumber
  \end{align}
  where  $\f_r(R_0,s)$ is the total probability flux into the spherical target:
  \begin{equation}
   \f_r(R_0,s)=\Omega_d R_1^{d-1} \J_r(R_1,s|R_0), 
     \end{equation}
 with $\Omega_d$ the solid angle of the $d$-dimensional sphere and 
 \begin{equation}
  \J_r(R_1,s|R_0)=\kappa_0 \p_r(R_1,s|R_0).
  \end{equation}
  We can identity $\f_r(R_0,s)$ as the Laplace transform of the FPT for adsorption. Setting $R_0=R_1$ and rearranging determines $\widetilde{\rho}_r(R,s|R_1)$ such that
 \begin{align}
 \label{2DLTrho}
 \widetilde{\rho}_r(R,s|R_0)&= \p_r(R,s|R_0)\\
 &\quad + \Lambda_r(R_1,s|R_0)\p_r(R,s|R_1),\nonumber 
 \end{align}
 with
  \begin{equation}
\label{2Dlambo}
\Lambda_r(R_1,s|R_0)=\frac{\pi_d\  \wphi(s )\f_r(R_0,s)}{1-  \pi_d  \wphi(s)\f_r(R_1,s)} .
\end{equation}
Similarly, equation (\ref{2DLT}b) becomes
 \begin{align}
 \widetilde{\calF}_r(R_0,s)&=\pi_b  \wphi(s)  \f_r(R_0,s)  \\
  &\quad  +\pi_d  \wphi(s)   \widetilde{\calF}_r(R_1,s) \f_r(R_0,s).\nonumber 
 \end{align}
 Again setting $R_0=R_1$ and rearranging determines $ \widetilde{\calF}_r(R_1,s)$ such that
 \begin{eqnarray}
 \label{2DLTcalF}
  \widetilde{\calF}_r(R_0,s)=\frac{\pi_b  \wphi(s )\f_r(R_0,s)}{1-  \pi_d   \wphi(s)\f_r(R_1,s)} .
 \end{eqnarray}

The
 probability density $\p(R,s|R_0)$ without desorption satisfies the Robin BVP
\begin{subequations}
\label{sph}
\begin{align}
  &D\frac{\partial^2\p_r(R,s|R_0)}{\partial R^2} + D\frac{d - 1}{R}\frac{\partial \p_r(R,s|R_0)}{\partial R}\nonumber \\
  &\quad -s\p_r(R,s|R_0) 
   =- \Gamma_0 \delta(R-R_0) \\
   &\hspace{4cm}- \Gamma_r \S_r(R_0,s)] \delta(R-R_r) ,   \nonumber \\
   &D\frac{\partial \p_r(R,s|R_0)}{\partial R}=\kappa_0 \p_r(R,s|R_0) ,\quad R=R_1.
\end{align}
\end{subequations}
We have set $\Gamma_0=1/(\Omega_dR_0^{d-1})$ and $\Gamma_r=1/(\Omega_dR_r^{d-1})$. Equations of the form (\ref{sph}) can be solved in terms of modified Bessel functions \cite{Redner01}. The general solution is
  \begin{align}
\label{qir}
    \p_r(R, s|R_0)& = A_r(s)\rho^\nu K_\nu(\alpha R) + G(R, \alpha| R_0) \\
    &\quad +r\S_r(R_0,s)G(R, \alpha | R_r), \ R_1<R,\nonumber
\end{align}
where $\nu = 1 - d/2$, $\alpha=\sqrt{[r+s]/D}$,  and $K_{\nu}$ is a modified Bessel function of the second kind.
The first term on the right-hand side of Eq. (\ref{qir}) is the solution to the homogeneous version of Eq. (\ref{sph}) and $G$ is the modified Helmholtz Green's function in the case of a totally absorbing surface $\partial \calU$ \cite{Redner01}:
\begin{align}
\label{GGs}
   G(R, s| R_0) &= \frac{ (RR_0)^\nu }{D\Omega_d}\bigg [\frac{[I_{\nu}(\alpha R_<)K_{\nu}(\alpha R_1)}{K_{\nu}(\alpha R_1)} \\
  &\qquad -\frac{I_{\nu}(\alpha R_1)K_{\nu}(\alpha R_<)}{K_{\nu}(\alpha R_>)} \bigg ],\nonumber 
\end{align}
where $R_< = \min{(R,R_0)}$, $R_> = \max{(R, R_0)}$, and $I_{\nu}$ is a modified Bessel function of the first kind.
The unknown coefficient $A_r(s)$ is determined from the boundary condition (\ref{sph}b):
\begin{align}
  \kappa_0 A_r F_{\alpha}(R_1) &=DA_r   F_{\alpha}'(R_1)+D\left . \frac{d}{dR}G(R,\alpha|R_0) \right |_{R=R_1} \nonumber \\
  &\quad +Dr\S_r(R_0,s)\left . \frac{d}{dR}G(R,\alpha|R_r) \right |_{R=R_1},
\label{CB3}
\end{align}
with
\begin{eqnarray}
\label{boll}
\left . D\frac{d}{dR}G(R,\alpha|R_0) \right |_{R=R_1}= \frac{1}{\Omega_dR_1^{d-1}} \frac{F_{\alpha}(R_0)}{F_{\alpha}(R_1)}.
\end{eqnarray}
We have set
\begin{equation}
\label{FF}
F_{\alpha}(R)=R^{\nu}K_{\nu}(\alpha R)
\end{equation}
so that
\begin{equation}
 F_{\alpha}'(R)=\nu R^{\nu-1} K_\nu(\alpha R) +\alpha R^{\nu}K'_{\nu}(\alpha R).
\end{equation}
Rearranging (\ref{CB3}) shows that 
\begin{align}
\label{ars}
&A_r(s)\\
&=\frac{D\bigg [\partial_RG(R_1,\alpha|R_0)  
+r\S_r(R_0,s)\partial_RG(R_1,\alpha|R_r)\bigg ]}{\kappa_0F_{\alpha}(R_1)-DF_{\alpha}'(R_1)}.\nonumber 
  \end{align}

\begin{figure}[t!]
\centering
\includegraphics[width=8.5cm]{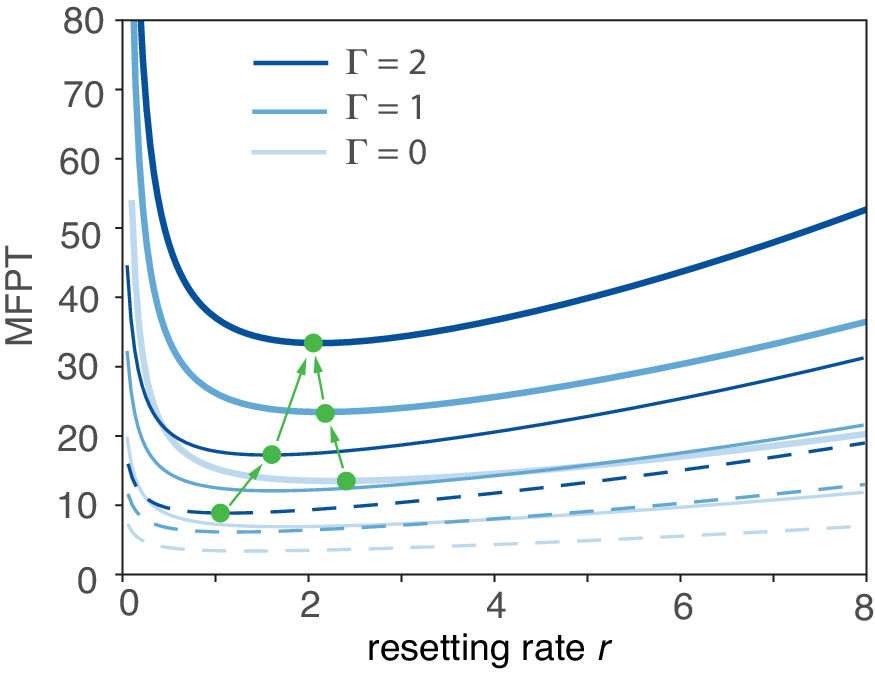} 
\caption{Effect of spatial dimension $d$ on the MFPT $\calT_r(x_0)$ for absorption by a spherical target of radius $R_1=1$. Plots of  $\calT_r(x_0)$ versus the resetting rate $r$ for different values of $\Gamma=\pi_d/\pi_b$ and $d=3$ (thick curves), $d=2$ (thin curves) and $d=1$ (dashed curves). Other parameters are $\langle \tau\rangle =0$, $R_0=R_r=2$, $\kappa_0=1$ and $D=1$. (Filled circles indicate optimal resetting rates.)}
\label{fig10}
\end{figure}

Hence, the full solution of the Laplace transformed probability density with resetting is
  \begin{align}
     \label{Rr}
   \p_r(R, s|R_0)& =  \p_0(R,r+s|R_0) \\
   &\quad +r\S_r(R_0,s) \p_0(R,r+s|R_r),\nonumber
\end{align}
where $\p_0$ is the corresponding solution without resetting,
\begin{align}
\label{pr0}
    \p_0(R, s|R_0) =A_0(s) R^\nu K_\nu(\sqrt{s/D} R) + G(R, \sqrt{s/D}| R_0).\nonumber \\
    \end{align}
Multiplying both sides of equation (\ref{Rr}) by $\Omega_dR^{d-1}$ and integrating with respect to $R$ implies that
\begin{align}
\label{Qr0}
   \S_r(R_0, R_r,s) &= \S_0(R_0,r+s)\\
   &\quad +r\S_r(R_0,s)]\S_0(R_r,r+s),\nonumber 
\end{align}
where $\S_0$ is the corresponding survival probability without resetting. Rearranging this equation yields the analog of Eq. (\ref{dSren0}):
\begin{equation}
\label{Qrel}
\S_r(R_0,s)=\frac{\S_0(R_0,r+s)}{1-r\S_0(R_r,r+s)}.
\end{equation}
The final step is to calculate the FPT densities with and without desorption. First, from Eqs. (\ref{boll}), (\ref{ars}) and (\ref{pr0}) we have
\begin{align}
\f_0(R_0,s)&=\kappa_0 \Omega_d R_1^{d-1} \p_0(R_1,s|R_0)\\
&= \left [   \frac{\kappa_0F_{\alpha}(R_0)}{\kappa_0F_{\alpha}(R_1)-DF_{\alpha}'(R_1)}\right ]_{\alpha =\sqrt{s/D}}.\nonumber
     \end{align}
     Second, from Eq. (\ref{Qrel}) we deduce that
   \begin{align}
 \f_r(R_0,s)
 &=\frac{ r\f_0(R_r,r+s)+s\f_0(R_0,r+s) }{s+r\f_0(R_r,r+s)} \\
 &=\frac{\kappa_0[rF_{\alpha}(R_r)+sF_{\alpha}(R_0)]}{s[\kappa_0F_{\alpha}(R_1)-DF_{\alpha}'(R_1)]+r\kappa_0F_{\alpha}(R_r)}. \nonumber
  \end{align}
 Expanding the right-hand side in powers of $s$ shows that
 \begin{align}
& T_r(x_0)\\
&=\frac{1}{r}\left [   \frac{\kappa_0F_{\alpha}(R_1)-DF_{\alpha}'(R_1)}{\kappa_0F_{\alpha}(R_r)}-\frac{F_{\alpha}(R_0)}{F_{\alpha}(R_r)}\right ]_{\alpha =\sqrt{r/D}}.\nonumber
\end{align}

In Fig. \ref{fig10} we show example plots of the MFPT for absorption, which is given by
\begin{align}
 \calT_r(R_0)=T_r(R_0)+  \langle \tau \rangle+\frac{\pi_d}{\pi_b}\bigg [T_r(R_1)+ \langle \tau \rangle\bigg ].
\label{dMFPT}
 \end{align}
 We plot $ \calT_r(R_0)$ as a function of $r$ for various levels of desorption and with $R_r=R_0$. It can be seen that the MFPT increases significantly with the dimension $d$. Moreover the optimal resetting rate increases with $d$ and decreases with $\Gamma=\pi_d/\pi_b$.

\setcounter{equation}{0}
\section{Encounter-based model of absorption}

So far we have assumed that when the searcher attaches to the target boundary $\partial \calU$, it spends a random waiting time $\tau$ at the boundary after which it either detaches with probability $\pi_d$ or enters the interior $\calU$ with probability $\pi_b=1-\pi_d$. In this section we consider one mechanism for generating a non-exponential waiting time density. We assume that the searcher detaches at a constant rate $\gamma_0$ as before. However, now the probability of final absorption depends on the amount of accumulated time that the searcher has spent attached to the boundary over successive visits to the target. This is a static version of the more familiar encounter-based model of diffusion-mediated adsorption, which assumes that the probability of adsorption depends upon the amount of particle-surface contact time prior to binding \cite{Grebenkov20,Grebenkov22,Bressloff22,Bressloff22a,Grebenkov24}. The amount of contact time is determined by a Brownian functional known as the boundary local time $\ell(t)$ \cite{Ito65,McKean75}. An adsorption event is identified as the first time that the local time crosses a randomly generated threshold $\widehat{\ell}$. This yields the stopping condition $T=\inf\{t>0, \ell(t) >\widehat{\ell} \}$. Different models of adsorption then correspond to different choices of the random threshold probability density $\psi(\ell)$. If $\psi(\ell)=\gamma_0\e^{-\gamma_0\ell}$, then the probability of adsorption over an infinitesimal local time increment is independent of the accumulated local time and we have Markovian adsorption with constant rate $\gamma_0$. On the other hand, a non-exponential distribution represents non-Markovian adsorption. We will apply the analogous encounter-based framework to absorption, by adapting a previous study that considered the absorption of an RTP at a partially absorbing sticky boundary \cite{Bressloff23}. (We could also consider a joint encounter-based model of adsorption and absorption but, for simplicity, we focus on the latter here.)

For the sake of illustration, we return to the example of diffusion with resetting on the half-line, see Sect. II. Define the occupation time $A(t)$ as the amount of time that the particle spends attached to the end $x=0$ over the time interval $(0,t)$. That is,
\begin{equation}
A(t)=\int_0^tq(t)d\tau,
\end{equation}
where $q(\tau)$ is the probability that the particle is in the bound state at time $\tau$. We then assume that the particle is absorbed as soon as the occupation time exceeds a randomly generated threshold $\widehat{a}$. The stopping time for absorption is thus
\begin{equation}
{\mathcal T}=\inf\{t>0:\ A(t)>\widehat{a}\},\quad \P[\widehat{a}>a]\equiv \Psi(a).
\end{equation}
Since $A(t)$ is a nondecreasing process, the condition $t<{\mathcal T}$ is equivalent to the condition $A(t)<\widehat{a}$.
It is also convenient to introduce the discrete variable $N(t)$ with $N(t)=1$ if the particle is freely diffusing and $N(t)=0$ if it is in the bound state at $x=0$.

The next step is to introduce the joint probability density or occupation time propagator
\begin{align}
&P(x,a,t|x_0)\nonumber \\
&=\bigg \langle \delta_{N(t)=1}\delta(x-X(t))\delta(a-A(t))\bigg \rangle_{X(0)=x_0},
\end{align}
where $\langle \cdot\rangle$ denotes expectation with respect to all sample paths $X(t)$ without absorption, that is, for a particle undergoing Brownian motion on the half-line combined with adsorption, desorption and resetting to $x_0$. Similarly. set
\begin{equation}
Q(a,t|x_0)= \bigg \langle \delta_{N(t)=0}\delta(a-A(t))\bigg \rangle_{X(0)=x_0}.
\end{equation}
It can be shown that
\begin{align}
\rho(x,t|x_0)&=\int_0^{\infty}P(x,a,t|x_0)dx,\\ q(x_0,t)&=\int_0^{\infty} Q(a,t|x_0)da,
\end{align}
where $\rho,q$ are the solutions to Eqs. (\ref{1D}) for $\overline{\gamma}_0=0$. In order to incorporate absorption, we define the marginal distributions
\begin{subequations}
\begin{align}
  \rho^{\Psi}(x,t|x_0)&= \bigg \langle \delta_{N(t),1} \delta(x-X(t))\Psi(A(t))\bigg \rangle_{X(0)=x_0}\nonumber \\
&=\int_0^{\infty}\Psi(a) P(x,a,t|x_0)da, \\
q^{\psi}(x_0,t)&=\bigg \langle \delta_{N(t),0}\psi(A(t))\bigg \rangle_{X(0)=x_0}  \nonumber \\
&=\int_0^{\infty}\Psi(a) Q(a,t|x_0)da,\\
 \nu^{\psi}(x_0,t)&=\bigg \langle \delta_{N(t),0}\psi(A(t))\bigg \rangle_{X(0)=x_0}  \nonumber \\
 &=\int_0^{\infty}\psi(a) Q(a,t|x_0)da,
\end{align}
\end{subequations}
with $\psi(a) =-\Psi'(a)$ the probability density for the occupation time threshold. Following similar arguments to Ref. \cite{Bressloff23}
it can be proved that the marginal distributions are related according to the equations
\begin{subequations}
\label{1DEB}
\begin{align}
 \frac{\partial \rho^{\psi}(x,t|x_0)}{\partial t}&=D\frac{\partial^2 \rho^{\Psi}(x,t|x_0)}{\partial x^2} -r\rho^{\psi}(x,t|x_0)\nonumber \\
 &\quad +r \calS^{\Psi}(x_0,t)\delta(x-x_0),\quad x>0,\\
D\frac{\partial \rho^{\Psi}(0,t|x_0)}{\partial x}&=\kappa_0\rho^{\Psi}(0,t|x_0) -\gamma_0 q^{\Psi}(t), 
\end{align}
with
\begin{equation}
\frac{dq^{\Psi}(x_0,t) }{dt}=\kappa_0\rho^{\Psi}(0,t|x_0) - \gamma_0 q^{\Psi}(x_0,t)-\nu^{\psi}(x_0,t).
\end{equation}
\end{subequations}

For a general occupation time threshold distribution $\Psi$, we do not have a closed system of equations for $\rho^{\Psi},q^{\Psi}$. However, in the particular case of the exponential distribution $\Psi(a)=\e^{-\overline{\gamma}_0a}$, we have $\psi(a)=\overline{\gamma}_0 \Psi(a)$ so that Eqs. (\ref{1DEB}) reduce to Eqs. (\ref{1D})
Hence, we have the identities
\begin{subequations}
\begin{align}
\calP(x,z,t|x_0)&\equiv \int_0^{\infty} \e^{-z a} P(x,a,t|x_0)da\nonumber \\
&=p(x,t|x_0)_{\kappa_0=z},\\
\calQ(z,t|x_0)&\equiv \int_0^{\infty} \e^{-z a} Q(a,t|x_0)da=q(x_0,t)_{\kappa_0=z}.
\end{align}
\end{subequations}
Assuming that the Laplace transforms can be inverted with respect to $z$, the solutions for a general distribution $\Psi$ are
 \begin{align}
  \label{poo}
  \rho^{\Psi}(x,t|x_0) &= \int_0^{\infty} \Psi(\ell){\mathbb L}_{a}^{-1}[\calP(x,z,t|x_0)]da,\\
   q^{\Psi}(x_0,t) &= \int_0^{\infty} \Psi(\ell){\mathbb L}_{a}^{-1}[\calQ(z,t|x_0)]da .
  \end{align}
 
 Working in Laplace space we can combine Eqs. (\ref{p1D})--(\ref{Sren}) to deduce that
   \begin{align}
  \q(x_0, s) &= \frac{\kappa_0}{s+\gamma_0+\overline{\gamma}_0} \frac{ 1}{1-r\widetilde{\calS}(x_0,r,s)}  \frac{\e^{-\alpha x_0}}{r+s+\alpha \kappa(s)}\nonumber \\
  &= \frac{\kappa_0}{s+\gamma_0+\overline{\gamma}_0}  \frac{\alpha(r+s)\e^{-\alpha x_0}}{s(r+s)+\alpha \kappa(s)[s+r\e^{-\alpha x_0}]}.
\end{align}
Substituting for $\kappa(s)$ using Eq. (\ref{ks}) and setting $\overline{\gamma}_0=z$ gives (after some algebra)
 \begin{align}
\widetilde{\calQ}(z,s|x_0)&
  &=  \frac{\kappa_0(r+s)\alpha\e^{-\alpha x_0}}{s(r+s)+\alpha \kappa_0[s+r\e^{-\alpha x_0}]}\frac{1}{z+\Theta(r,s)},
\end{align}
 where
 \begin{equation}
 \Theta(r,s)=s\left [1+\frac{\gamma_0(r+s)}{s(r+s)+\kappa_0\alpha[s+r\e^{-\alpha x_0}]}\right ].
 \label{Thet}
 \end{equation}
  Inverting with respect to $z$ yields the propagator
   \begin{align}
\widetilde{Q}(a,s|x_0)&
  &=  \frac{\kappa_0(r+s)\alpha\e^{-\alpha x_0}}{s(r+s)+\alpha \kappa_0[s+r\e^{-\alpha x_0}]}\e^{-a\Theta(r,s)},
\end{align}
which implies that
\begin{align}
\widetilde{\nu}^{\psi}(x_0,s)&= \frac{\kappa_0(r+s)\alpha\e^{-\alpha x_0}}{s(r+s)+\alpha \kappa_0[s+r\e^{-\alpha x_0}]}\widetilde{\psi}(\Theta(r,s))\nonumber \\
&=\frac{\widetilde{\psi}(\Theta(r,s))}{\widetilde{\Psi}(\Theta(r,s))}\widetilde{q}^{\Psi}(x_0,s).
\end{align}

It follows that in the time domain the constant rate of absorption is replaced by an effective time-dependent kernel $\gamma_b(t)$ such that
\begin{equation}
\nu^{\psi}(x_0,t)=\int_0^t {\gamma}_b(\tau) q^{\Psi}(t-\tau)d\tau,
\end{equation}
with
\begin{equation}
\widetilde{\gamma}_b(s) =\frac{\widetilde{\psi}(\Theta(r,s))}{\widetilde{\Psi}(\Theta(r,s))}.
\label{gamb}
\end{equation}
We highlight a few properties of the delay kernel $\gamma_b(\tau)$. First,
\begin{align}
\int_0^{\infty} \gamma_b(\tau)d\tau&\equiv \widetilde{\gamma}_b(0)=\frac{\widetilde{\psi}(\Theta(r,0))}{\widetilde{\Psi}(\Theta(r,0))}\nonumber \\
&=\frac{\widetilde{\psi}(0)}{\widetilde{\Psi}(0)}=-\frac{1}{\widetilde{\psi}'(0)}=\langle a \rangle^{-1}.
 \end{align}
Here $\langle a\rangle$ is the mean occupation time threshold (assuming it exists). Second,
\begin{align}
\int_0^{\infty}\tau \gamma_b(\tau)d\tau&\equiv -\widetilde{\gamma}_b'(0)\\
&=-\left [\frac{\widetilde{\psi'}(0)}{\widetilde{\Psi}(0)}-\frac{\widetilde{\Psi}'(0)}{\widetilde{\Psi}(0)^2}\right ]\partial_s\Theta(r,0)\nonumber \\
&=\left [1-\frac{\langle a^2\rangle}{2\langle a\rangle^2}\right ]\bigg [1+\frac{\gamma_0\e^{\sqrt{r/D} x_0}}{\kappa_0\sqrt{r/D}]}\bigg ].\nonumber 
 \end{align}
 Since the mean of $\gamma_b(\tau)$ is negative for densities $\psi(a)$ whose low-order moments satisfy the inequality$\langle a^2\rangle >2\langle a\rangle^2$, it follows that the effective kernel $\gamma_b(\tau)$ may not be positive for all $\tau \in [0,\infty)$. 
 
 For the sake of illustration, consider one well-known example of a non-exponential density with finite moments, namely, the gamma distribution:
\begin{equation}
\label{phigam}
\psi(a)=\frac{\gamma(\overline{\gamma}_0 a)^{\mu-1}\e^{-\overline{\gamma}_0 a}}{\Gamma(\mu)},\quad \mu >0,
\end{equation}
where $\Gamma(\mu)$ is the gamma function. If $\mu=1$, then we recover the exponential distribution $\psi(a)=\overline{\gamma}_0\e^{-\overline{\gamma}_0a}$. From the definition of the gamma distribution, we see that the probability of small values of $a$ can be decreased relative to an exponential distribution by taking $\mu >1$. This could represent a bound state that is initially relatively stable, but becomes more unstable  as $a$ increases. On the other hand, the probability of small values of $a$  is increased when $\mu < 1$ so that the bound state is initially more unstable.
The corresponding Laplace transform is
\begin{equation}
\widetilde{\psi}(z)=\left (\frac{\overline{\gamma}_0}{z+\overline{\gamma}_0}\right )^{\mu},
\end{equation}
and the moments are
\begin{equation}
\langle a^n\rangle =\left .\left (-\frac{d}{dz}\right )^n\widetilde{\psi}(z)\right |_{z=0}=\frac{\mu (\mu+1)\ldots (\mu+n-1)}{\overline{\gamma}_0^{n}}.
\end{equation}
In particular, $\langle a^2\rangle =\langle a\rangle^2 (1+1/\mu)$, which implies that the mean of $\gamma_b(\tau)$ is positive provided that $\mu \geq 1$.

Finally, we can incorporate the above construction into the renewal equation by considering the equivalent waiting time density $\phi(\tau)$ and associated splitting probabilities. Suppose that the particle binds to the target (is adsorbed) at a time $t_a$, say. Let $\sigma(\tau)$ be the probability that it is still bound at time $t_a+\tau$. (The probability $\sigma(\tau)$ is distinct from the probability $q^{\Psi}(t)$, since the former is conditioned on adsorption occurring at time $\tau=0$.) We have
 \begin{equation}
 \frac{d\sigma}{d\tau}=-\gamma_0 \sigma(\tau)-\int_0^t\gamma_b(\tau)\sigma(t-\tau)d\tau,\quad \sigma(0)=1.
 \end{equation}
Laplace transforming this equation under the given initial condition yields
\begin{equation}
\widetilde{\sigma}(s)=\frac{1}{s+\gamma_0+\widetilde{\gamma}_b(s)}.
\end{equation}
Identifying the waiting time density for either desorption or absorption to occur as $\phi(\tau)=-d\sigma /d\tau $, we see that
 \begin{equation}
\wphi(s)=\frac{ \gamma_0+\widetilde{\gamma}_b(s)}{s+ \gamma_0+\widetilde{\gamma}_b(s)}.
\end{equation}
In addition, denoting the splitting probabilities for desorption and absorption by $\pi_d$ and $\pi_b$, respectively, we have
 \begin{equation}
 \pi_d =\frac{ \gamma_0}{\gamma_0+\widetilde{\gamma}_b(0)},\quad \pi_b=\frac{\widetilde{\gamma}_b(0)}{ \gamma_0+\widetilde{\gamma}_b(0)}.
 \end{equation}
One major consequence of the encounter-based model of absorption is that the effective waiting time density $\phi(\tau)$ depends on parameters of the diffusive search process, namely, the diffusivity $D$, the resetting rate $r$ and position $x_0$, and the adsorption rate $\kappa_0$. This follows from Eq. (\ref{Thet}).

\section{Discussion}  

In this paper we developed a general mathematical framework for analyzing the combined effects of target-searcher interactions and stochastic resetting on the diffusive search for resources contained within a partially accessible target $\calU\subset \R^d$. By analogy with diffusion-mediated surface reactions in physical chemistry, we modeled interactions with the target surface $\partial \calU$ in terms of particle adsorption, desorption and absorption. First, we assumed that when the searcher binds (adsorbs) to the surface $\partial \calU$, it spends a random waiting time $\tau$ with associated density $\phi(\tau)$ attached to the surface, after which it either gains access to the resources (absorption) or detaches and continues its search process (desorption). Second, we assumed that the target surface $\partial \calU$ is partially adsorbing, that is, there is a nonzero probability that the particle reflects off the surface each time it encounters the target.

\begin{figure}[b!]
\centering
\includegraphics[width=8cm]{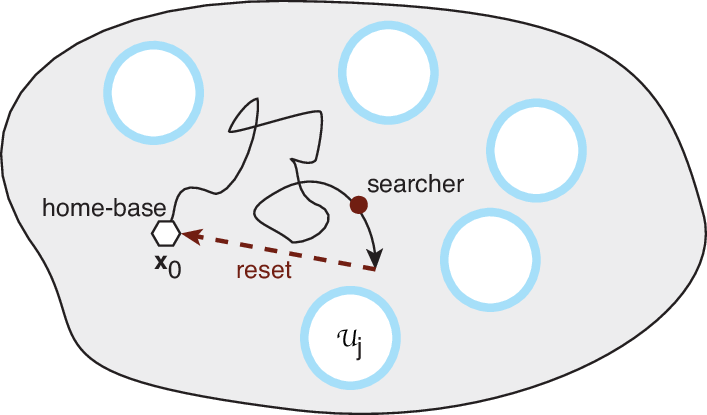} 
\caption{Search for resources in a domain with multiple partially accessible targets $\calU_j$, $j=1,\ldots,N$..}
\label{fig11}
\end{figure}

One of the main results of our work is the reformulation of a search process with stochastic resetting and partially accessible targets in terms of a pair of renewal equations that relate the probability density $\rho_r(\x,t|\x_0)$ and FPT density $\calF_r(\x_0,t)$ in the presence of absorption to the corresponding quantities $p_r(\x,t|\x_0)$ and $f_r(\x_0,t)$ for irreversible adsorption. The renewal equations have several advantages over standard PDEs. First, they provide a direct method for incorporating arbitrary non-exponential waiting time densities, which correspond to non-Markovian models of absorption and desorption. Second, they express important quantities such as the FPT moments in terms of the statistics of $\phi(\tau)$ and the number of desorptions. Third, renewal equations can be constructed irrespective of whether the searcher immediately returns to its home base after each desorption event or continues from the point on the surface $\partial \calU$ where it detaches. However, the latter case requires two major modifications: (i) Distinguishing between the reset point and the initial position following each desorption event; (ii) Integrating with respect to the points $\y \in \partial \calU$ where the particle detaches. (The latter isn't required in 1D.)

In our single-target model we assumed that if the particle fails to enter the domain $\calU$ and desorbs, then it continues the random search process as before. This is based on  the additional assumptions that (i) the searcher does not know whether there are additional targets within the search domain and (ii) whenever the searcher returns to the same target it interacts in the same way as previous visits. A major area of future work is to extend the theory to multiple partially accessible targets, as illustrated in Fig. \ref{fig11}.
A number of issues then arise. First, in the case of a single target it is clearly advantageous to restart the search process at the target rather than resetting to $\x_0$. However, this is no longer necessarily the case when there are multiple targets. The most efficient search protocol will depend on how the targets are distributed across the search domain and the location of the home base. This then raises a second issue, namely, is it possible to minimize the expected time to gain access to resources by varying features such a the resetting rate, the reset location and the distribution of targets? Third, one could include the additional constraint that the particle does not attach to a target it has previously failed to enter. However, this introduces memory into the system. Finally, in order to analyze the multi-target case in $\R^d$, $d>1$, it will be necessary to deal with the spatial integrals appearing in the multi-target generalization of the renewal Eqs. (\ref{2Dren}). As highlighted in Appendix A, this is a non-trivial problem. However, it might be possible to avoid such technicalities by working with a small-target limit and using matched asymptotics methods \cite{Bressloff24}.

\setcounter{equation}{0}
\renewcommand{\theequation}{A.\arabic{equation}}
\section*{Appendix A: Spectral decomposition of the renewal equations}

In this appendix we show how the spectral decomposition of the solution to the Robin BVP (\ref{PlocLT}) in terms of the eigenfunctions of a Dirchlet-to-Neumann (D-to-N) operator on $\partial \calU$ \cite{Grebenkov19a} can be used to solve the higher-dimensional renewal equations (\ref{2DLT}).
The general solution of Eq. (\ref{PlocLT}a) with the effective boundary condition $\p_0(\y,s|\x_0)=h(\x_0,s)$ for all $\y \in \calU$ is
 \begin{align}
 \label{sool}
 \p_0(\x,s|\x_0)&= \calH(\x,s|\x_0)+G(\x,s|\x_0),\ \x \notin \overline{\calU}  \end{align}
 where 
 \begin{equation}
  \calH(\x,s|\x_0)=-D\int_{\partial \calU} \partial_{\sigma'} G(\x',s|\x)h(\x',s|\x_0)d\x' 
  \end{equation}
  for $\partial_{\sigma'}=\n_0\cdot {\bm \nabla}_{\x'}$, and $G $ is a modified Helmholtz Green's function:
  \begin{align}
 &D{\bm \nabla}^2 G(\x,s|\x')-sG(\x,s|\x')=-\delta(\x-\x'),\ \x,\x' \notin \overline{\calU},\nonumber \\
 & G(\y,s|\x') =0,\ \y\in \partial \calU  .
  \label{nabG}
 \end{align}
The unknown function $h$ is determined by substituting the solution (\ref{sool}) into Eq. (\ref{PlocLT}b):
\begin{align}
\label{fL}
&\L_s[h](\y,s|\x_0)+\frac{\kappa_0}{D}h(\y,s|\x_0)=-\partial_{\sigma}G(\y,s|\x_0)
\end{align}
for all $  \y \in \partial \calU$ where $\L_s$ is the D-to-N operator
 \begin{eqnarray}
\label{DtoN}
\L_s[F](\y) &=-D\partial_{\sigma}\int_{\partial \calU}\partial_{\sigma'}G(\x',s|\x)F(\x')d\x'
\end{eqnarray}
acting on the space $L^2(\partial \calU)$. 

When the surface $\partial \calU$ is bounded, the D-to-N operator $\L_s$ has a discrete spectrum. That is, there exist countable sets of eigenvalues $\mu_n(s)$ and eigenfunctions $v_n(\x,s)$ satisfying (for fixed $s$)
\begin{equation}
\label{eig}
\L_s v_n(\y,s)=\mu_n(s)v_n(\y,s),\quad \y\in \partial \calU
\end{equation}
It can be shown that the eigenvalues are non-negative and that the eigenfunctions form a complete orthonormal basis in $L^2(\partial \calU)$. We can now solve Eq. (\ref{fL}) by introducing an eigenfunction expansion of $h$,
\begin{equation}
\label{eig2}
h(\y,s|\x_0)=\sum_{m=0}^{\infty}h_m(\x_0,s) v_m(\y,s).
\end{equation}
Substituting Eq. (\ref{eig2})
into (\ref{fL}) and taking the inner product with the adjoint eigenfunction $v_n^*(\x,s)$ determines the coefficients $h_m$:
\begin{align}
&(\mu_n(s)+\kappa_0/D)h_m(\x_0,s) ={\mathcal V}^*_n(\x_0,s),
\label{hm}
\end{align}
with
\begin{equation}
\label{VVn}
{\mathcal V}_n(\x,s)=-D\int_{\partial \calU}v_n(\x',s) \partial_{\sigma'}G(\x',s|\x)d\x'.
\end{equation}
By construction we have the identities
\begin{equation}
\label{ids}
{\mathcal V}_n(\y,s)=v_n(\y),\ \partial_{\sigma} {\mathcal V}_n(\y,s)=\mu_n(s)v_n(\y,s),\quad \y \in \partial \calU.
\end{equation}
Note that the orthogonality condition 
\begin{equation}
\int_{\partial \calU} v_n^*(\x,s)v_m(\x,s)d\x=\delta_{m,n}
\end{equation}
means that $v_n^*$ and $v_m$ can each be taken to have dimensions of [Length]$^{-(d-1)/2}$.
Finally, substituting Eqs. (\ref{eig2}) and (\ref{hm}) into the solution (\ref{sool}) yields
\begin{align}
 \label{spec2}
\p_0(\x,s|\x_0)=G(\x,s|\x_0)+\frac{1}{D}\sum_{n=0}^{\infty} \frac{{\mathcal V}_n(\x,s){\mathcal V}^*_n(\x_0,s)}{\mu_n(s)+\kappa_0/D}.
\end{align}
It is convenient to rewrite the solution (\ref{spec2}) as \cite{Grebenkov23,Bressloff25}
\begin{equation}
\p_0(\x,s|\x_0)=\p_0(\x,s|\x_0)_{\kappa_0=0}-\frac{1}{D}\sum_{n=0}^{\infty} \frac{{\mathcal V}_n(\x,s){\mathcal V}^*_n(\x_0,s)}{\mu_n(s)[D\mu_n(s)/\kappa_0+1]},
\end{equation}
with $\partial_{\sigma}\p_0(\y,s|\x_0)_{\kappa_0=0}=0$ for $\y\in \partial \calU$.
Using the identities (\ref{ids}) we have
 \begin{align}
  \J_0(\y,s|\x_0)&=-D\partial_{\sigma} \p_r(\y,s|\x_0) \nonumber \\
  &\quad =\frac{\kappa_0}{D}\sum_{n=0}^{\infty} \frac{ v_n(\y,s){\mathcal V}^*_n(\x',s)}{\mu_n(s)+\kappa_0/D}
 \end{align}
 and
   \begin{align}
  \J_0(\y,s|\y')=\frac{\kappa_0}{D}\sum_{n=0}^{\infty} \frac{ v_n(\y,s)v^*_n(\y',s)}{\mu_n(s)+\kappa_0/D}
    \end{align}
  for all $\y,\y' \in \partial \calU$,
 
 We can now solve the renewal Eq. (\ref{2DLT}b) for the FPT density in terms of the D-to-N eigenfunctions. Eq. (\ref{peep}) implies that
  \begin{align}
\J_r(\y,s|\x_0)&=\J_0(\y,r+s|\x_0) \nonumber \\
&\quad +r\S_r(\x_0,s)\J_0(\y,r+s|\x_r),
\label{peep2}
\end{align}
with
 \begin{align}
 \S_r(\x_0,s)&=\frac{1-\f_0(\x_0,r+s) }{s+r\f_0(\x_r,r+s)}
  \end{align}
  Moreover,
  \begin{equation}
  \f_0(\x_0,s)=\int_{\partial \calU}\J_0(\y,s|\x_0)d\y.
  \end{equation}
  Consider the trial solution
  \begin{equation}
  \label{tri1}
  \widetilde{\calF}_r(\x_0,s)=\widetilde{\calG}_r(\x_0,s)+r\S_r(\x_0,s)\widetilde{\calG}_r(\x_r,s)
  \end{equation}
  with
   \begin{align}
   \label{tri2}
\widetilde{\calG}_r(\x,s)&= \pi_b\wphi(s) \f_0(\x,r+s) -\Gamma_r(s)\\
&\quad +\pi_d \wphi(s)  \int_{\partial \calU}  \widetilde{\calG}_r(\y,s) \J_0(\y,r+s|\x)d\y\nonumber
 \end{align}
 for $\x=\x_0,\x_r$. The unknown function $\Gamma(s)$ has to be determined self-consistently. That is, given the relation
 \begin{align}
&\widetilde{\calG}_r(\x_0,s)+r\S_r(\x_0)\widetilde{\calG}_r(\x_r,s)\nonumber \\
&= \pi_b\wphi(s) \f_r(\x_0,s)-(1+r\S_r(\x_0,s))\Gamma(s)\nonumber  \\
&\quad +\pi_d \wphi(s)  \int_{\partial \calU}  \widetilde{\calG}_r(\y,r,s) \J_r(\y,s|\x_0)d\y,\nonumber
 \end{align}
 the trial solution will satisfy Eq. (\ref{2DLT}b) provided that
 \begin{align}
 \label{Gs}
 &(1+r\S_r(\x_0))\Gamma_r(s)\\
 &=\pi_d \wphi(s) \widetilde{\calG}_r(\x_r,s) \int_{\partial \calU} r\S_r(\y,s) \J_r(\y,s|\x_0)d\y.  \nonumber 
  \end{align}
  This is an inhomogeneous equation for $\Gamma_r(s)$ as $\widetilde{\calG}_r(\x_r,s) $ depends on $\Gamma_r(s)$, see below.
  
 The final step is to solve the simplified renewal Eq. (\ref{tri2}). Following along similar lines to Ref. \cite{Grebenkov23,Bressloff25}, we perform a Neumann expansion of the right-hand side after setting
 \begin{equation}
\Gamma_r(\x,s)= \pi_b\wphi(s) \f_0(\x,r+s) -\Gamma_r(s).
\end{equation}
This gives
\begin{widetext} \begin{align}
\label{Neum2D}
 \widetilde{\calG}_r(\x,s)&=\Gamma_r(\x,s) +\pi_d \wphi(s) \int_{\partial \calU}\Gamma_r(\y,s)\J_0(\y,r+s|\x) d\y\nonumber \\
 &\quad +(\pi_d \wphi(s) )^2 \int_{\partial \calU}\Gamma_r(\y,s)\bigg (  \int_{\partial \calU}\J_0(\y,r+s|\y') \J_0(\y',r+s|\x) d\y' \bigg )d\y+\ldots  \bigg ]  \end{align}
     \end{widetext}
  Substituting the series expansions of $\J_0$ and using the orthonormality of the D-to-N eigenfunctions implies that $\calF$ is given by a geometric series
 \begin{align}
 \widetilde{\calG}_r(\x,s)&=\Gamma_r(\x,s) + \pi_d\wphi(s)  \frac{\kappa_0}{D}\sum_{n=0}^{\infty} \frac{\overline{v}_n(r,s){\mathcal V}^*_n(\x,r+s)}{\mu_n(r+s)+\kappa_0/D}
 \nonumber \\
 &\times \bigg [1+\frac{\kappa_0}{D}\frac{\pi_d \wphi(s)}{\mu_n(r+s)+\kappa_0/D}+\ldots\bigg ],
  \end{align}
  where
  \begin{equation}
  \overline{v}_n(r,s)=\int_{\partial \calU}\Gamma(\y,s) v_n(\y,r+s)d\y.
  \end{equation}
  We conclude that
     \begin{align}
     \label{FaF}
& \widetilde{\calG}_r(\x,s)=\Gamma_r(\x,s) \\
&\quad +\sum_{n\geq 0}\left [ \frac{\pi_b  \wphi(s)\lambda_n(r+s)}{1-\pi_d \wphi(s)  \lambda_n(r+s)} \right ] \overline{v}_n(r,s){\mathcal V}^*_n(\x,r+s), \nonumber \end{align}
 with
 \begin{equation} \lambda_n(s)= \frac{\kappa_0}{D}\frac{1}{\mu_n(s)+\kappa_0/D}.
\end{equation}
Finally, substituting Eq. (\ref{FaF}) into Eq. (\ref{Gs}) yields an inhomogeneous linear equation for the unknown function $\Gamma(s)$. The integral term in (\ref{Gs}) can be evaluated by setting
\begin{align}
& \int_{\partial \calU} \S_r(\y,s) \J_r(\y,s|\x_0)d\y\\
&=\frac{1}{s+r\f_0(\x_r,r+s)} \int_{\partial \calU}d\y \bigg [ 1-\f_0(\y,r+s)\bigg  ] \nonumber \\
&\quad \times  \bigg [\J_0(\y,r+s|\x_0)+r\S_r(\x_0,s)\J_0(\y,r+s|\x_r)\bigg ],\nonumber
\end{align}
performing eigenfunction expansions of $\f_0$ and $\J_0$, and using the orthonormality of the D-to-N eigenfunctions.

We conclude that solving the higher-dimensional renewal Eqs. (\ref{2DLT}) is possible in principle using the spectral decomposition of the associated D-to-N operator (\ref{DtoN}). However, the detailed calculations are non-trivial and, as far as we are aware, exact expressions for the eigenvalues and eigenfunctions are only known in a few special cases. In addition, the final expressions involve infinite sums that need to be truncated. Therefore, to what extent spectral methods provide a practical tool for solving the higher-dimensional renewal equations remains to be determined. Alternatively, one can restrict the analysis to problems with additional constraints such as spherical symmetry or develop perturbative approximations.


\vfill
 
\begin{thebibliography}{9}


\bibitem{Evans11a} M. R. Evans and S. N. Majumdar, Diffusion with stochastic resetting. {Phys. Rev. Lett.}, {\bf 106},160601 (2011).

\bibitem{Evans11b}  M. R. Evans and S. N. Majumdar, Diffusion with optimal resetting, {J. Phys. A Math. Theor.}, {\bf 44}, 435001 (2011).


\bibitem{Evans14}  M. R. Evans and S. N. Majumdar, Diffusion with resetting
in arbitrary spatial dimension. { J. Phys. A}, {\bf 47}, 285001 (2014).

\bibitem{Evans20}  M. R. Evans, S. N. Majumdar and G. Schehr, Stochastic resetting and applications. {J. Phys. A}, {\bf 53}, 193001 (2020).

 \bibitem{Nagar23} A. Nagar and S. Gupta. Stochastic resetting in interacting particle
systems: a review {J. Phys. A: Math. Theor.} {\bf 56} 283001 (58pp) (2023).

\bibitem{Pal20}  A. Pal, L. Kusmierz and S. Reuveni, Home-range search provides advantage under high uncertainty. {Phys. Rev. Res.}, {\bf 2}, 043174 (2020).

\bibitem{Evans13} J. Whitehouse, M. R. Evans and S. N. Majumdar, Effect of partial absorption on diffusion with resetting {Phys. Rev. E} {\bf 87} 022118 (2013).

\bibitem{Bressloff22b} P. C. Bressloff,  Diffusion-mediated surface reactions and stochastic resetting. {J. Phys. A} {\bf 55} 275002 (2022)

 \bibitem{Benk22} Benkhadaj Z and Grebenkov D S 2022 Encounter-based approach to diffusion
with resetting, {\em Phys. Rev. E} {\bf 106} 044121


\bibitem{Pal19} A. Pal, L. Kusmierz and S. Reuveni, Diffusion with stochastic resetting is invariant to return speed {Phys. Rev. E} {\bf 100} 040101 (2019)

\bibitem{Pal19a}  A. Pal, L. Kusmierz and S. Reuveni, Invariants of motion with stochastic resetting and spacetime coupled returns {New J. Phys.} {\bf 21} 113024 (2019)


\bibitem{Mendez19} A. Maso-Puigdellosas, D. Campos and V. Mendez, Transport properties of random walks under stochastic noninstantaneous resetting. Phys. Rev. E {\bf 100} 042104 (2019).

\bibitem{Bodrova20} A. S. Bodrova and I. M. Sokolov. Resetting processes with non-instantaneous return. Phys. Rev. E {\bf 101} 052130 (2020).


\bibitem{Bressloff20} P. C. Bressloff. Modeling active cellular transport as a directed search process with stochastic resetting and delays. J. Phys. A 53 355001 (2020).

 \bibitem{Evans19a}  M. R. Evans and S. N. Majumdar, Effects of refractory period on stochastic resetting {J. Phys. A: Math. Theor.}, {\bf 52}, 01LT01 (2019).

\bibitem{Mendez19a} A. Maso-Puigdellosas, D. Campos and V. Mendez. Stochastic movement subject to a reset-and-residence mechanism: transport properties and first arrival statistics. J. Stat. Mech. 033201 (2019).





\bibitem{Baret68} J. F. Baret, Kinetics of adsorption from a solution. Role of the diffusion
and of the adsorption-desorption antagonism {J. Phys. Chem.} {\bf 7} 2755-2758 (1968).


\bibitem{Adam87} Z. Adamczyk and J. Petlicki, Adsorption and desorption kinetics of molecules
and colloidal particles {J. Colloid Interface Sci.} {\bf 118} 20-49 (1987).


\bibitem{Adam87a} Z. Adamczyk, Nonequilibrium surface tension for mixed adsorption kinetics
{J. Colloid Interface Sci.} {\bf 120} 477-485 (1987).


\bibitem{Franses95} C. H. Chang and E. I. Franses,  Adsorption dynamics of surfactants at
the air/water interface: A critical review of mathematical models, data, and
mechanisms. {Colloids Surf. A } {\bf 100} 1-45 (1995).

\bibitem{Passerone96} L. Liggieri, F. Ravera and A.Passerone, A diffusion-based approach to mixed
adsorption kinetics {Colloids Surf. A} {\bf 114} 351-359 (1996).


\bibitem{Reuveni23}
Y. Scher, O. L.  Bonomo, A. Pal and S. Reuveni, Microscopic theory of
adsorption kinetics. {J. Chem. Phys.} {\bf 158} 094107 (2023)

\bibitem{Agmon84} N. Agmon, Diffusion with back reaction {J. Chem. Phys.} {\bf 81} 2811 (1984).



\bibitem{Agmon89} N. Agmon and G. H. Weiss, Theory of non-Markovian reversible dissociation
reactions. {J. Chem. Phys.} {\bf 91} 6937-6942 (1989).

\bibitem{Agmon90} N. Agmon and A.Szabo, Theory of reversible diffusion-influenced reactions
{J. Chem. Phys.} {\bf 92} 5270-5284 (1990).

\bibitem{Agmon93} N. Agmon, Competitive and noncompetitive reversible binding processes
{Phys. Rev. E} {\bf 47} 2415 (1993).

\bibitem{Gopich99} I. V. Gopich, K. M. Solntsev and N, Agmon, Excited-state reversible geminate
reaction. I. Two different lifetimes {J. Chem. Phys.} {\bf 110} 2164-2174 (1999)



\bibitem{Kim99} H. Kim and K. J. Shin, Exact solution of the reversible diffusion-influenced
reaction for an isolated pair in three dimensions {Phys. Rev. Lett.} {\bf 82} 1578 (1999).

\bibitem{Tachiya13} T. Pr\^ustel and M. Tachiya, Reversible diffusion-influenced reactions of an
isolated pair on some two dimensional surfaces {J. Chem. Phys. } {\bf 139} 194103 (2013).


\bibitem{Prustel13} T. Pr\^ustel and M. Meier-Schellersheim, Theory of reversible diffusion-influenced
reactions with non-Markovian dissociation in two space dimensions
{J. Chem. Phys.} {\bf 138} 104112 (2013).

\bibitem{Grebenkov19}  D. S. Grebenkov, Reversible reactions controlled by surface diffusion on a
sphere {J. Chem. Phys.} {\bf 151} 154103 (2019).

\bibitem{Grebenkov23} D. S. Grebenkov, Diffusion-controlled reactions with non-Markovian
binding/unbinding kinetics {J. Chem. Phys.} {\bf 158} 214111 (2023).

\bibitem{Bressloff25} P. C. Bressloff. Diffusion-mediated adsorption versus absorption at partially reactive targets: a renewal approach. Preprint http://arxiv.org/abs/2503.01308 (2025)


\bibitem{Grebenkov19a} D. S. Grebenkov, Spectral theory of imperfect diffusion-controlled reactions
on heterogeneous catalytic surfaces {J. Chem. Phys.} {\bf 151} 10410 (2019)8.

\bibitem{Redner01}
S. Redner {\em A Guide to First-Passage Processes}.
\newblock Cambridge University Press, Cambridge, UK (2001)

\bibitem{Grebenkov20} D. S. Grebenkov, {Paradigm shift in diffusion-mediated surface phenomena.} {Phys. Rev. Lett.} {\bf 125} 078102 (2020).



\bibitem{Grebenkov22} D. S. Grebenkov, {An encounter-based approach for restricted diffusion with a gradient drift.}  {J. Phys. A.} {\bf 55} 045203 (2022)

\bibitem{Bressloff22} P. C. Bressloff,  Diffusion-mediated absorption by partially reactive targets: Brownian functionals and generalized propagators. {J. Phys. A.} {\bf 55} 205001 (2022)

\bibitem{Bressloff22a} P. C. Bressloff,  Spectral theory of diffusion in partially absorbing media. {Proc. R. Soc. A} {\bf 478} 20220319 (2022)



\bibitem{Grebenkov24} D. S. Grebenkov, Encounter-based approach to target search problems
In {\em Target Search Problems}  Springer Nature Switzerland pp. 77-105 (2024)

 \bibitem{Ito65} K. Ito and H. P. McKean, {\em Diffusion Processes and Their Sample Paths} Springer-Verlag (1965).
Berlin
 \bibitem{McKean75} H. P. McKean, {Brownian local time.} {\em Adv. Math.} {\bf 15} 91-111 (1975).



\bibitem{Bressloff23} P. C. Bressloff, Encounter-based model of a run-and-tumble particle II: absorption at sticky boundaries. J. Stat. Mech. {\bf 043208} (2023).

\bibitem{Bressloff24} P. C. Bressloff, Cellular diffusion processes in singularly perturbed domains,
J.f Math. Biol. {\bf 89}, 58 (2024).
\end{thebibliography}
\end{document}